\documentclass[aps,twocolumn,prc,floatfix,showpacs, preprintnumbers]{revtex4-1}

\usepackage{graphicx}
\usepackage{amssymb}
\usepackage{amsmath}
\usepackage{mathtools}
\usepackage[dvipsnames,usenames]{xcolor}
\usepackage{multirow}

\newcommand{\be}{\begin{equation}}
\newcommand{\ee}{\end{equation}}
\newcommand{\bs}[1]{\ensuremath{\boldsymbol{#1}}}

\newcommand{\bra}{\langle}
\newcommand{\ket}{\rangle}

\begin{document}

\preprint{LA-UR-22-23223}

\title{Isospin-symmetry implications for nuclear two-body distributions
and short-range correlations}
\author{ {Ronen} Weiss$^{\, {\rm a} }$,
{Alessandro} Lovato $^{\, {\rm b,c,d} }$,
{R. B.} Wiringa $^{\, {\rm b} }$
}
\affiliation{
$^{\,{\rm a}}$\mbox{Theoretical Division, Los Alamos National Laboratory, Los Alamos, New Mexico 87545, USA}\\
$^{\,{\rm b}}$\mbox{Physics Division, Argonne National Laboratory, Argonne, Illinois 60439, USA}\\
$^{\,{\rm c}}$\mbox{Computational Science Division, Argonne National Laboratory, Argonne, Illinois 60439, USA}\\
$^{\,{\rm d}}$\mbox{INFN-TIFPA Trento Institute of Fundamental Physics and Applications, Via Sommarive, 14, 38123 {Trento}, Italy}\\
}
\date{\today}

\begin{abstract}
We study the implications of isospin symmetry in nuclear
systems on two-body distributions and derive relations between different densities
or momentum distributions of a given nucleus or of different nuclei in the
same isospin multiplet. A connection between neutrinoless double beta decay transition
of mirror nuclei and two-body densities is also obtained.
The relations are numerically verified by means of quantum Monte Carlo calculations.
We discuss the relevance of these results to the analysis of nuclear short-range
correlations and also show that isospin symmetry allows us to extract
information about the spectator nucleons when a short-range correlated pair is formed.
\end{abstract}

\maketitle

\section{Introduction}

Although the proton has a positive charge, and the neutron is neutral, they are very similar particles, with nearly degenerate masses. Back in 1932, Heisenberg argued that the proton and neutron can be regarded as different isospin states of the same underlying nucleon particle~\cite{Heisenberg:1932dw}. Specifically, the protons and neutrons are considered to form an isospin doublet, with total isospin $T=1/2$, and isospin projection $t_z=1/2$ and $t_z=-1/2$ for the proton and the neutron, respectively.

The proton-proton ($pp$), neutron-neutron ($nn$), and proton-neutron ($pn$) interactions in a given total isospin channel are approximately equal, as the nuclear Hamiltonian commutes with the total isospin operator, up to small corrections due to the Coulomb force and other small isospin-breaking terms. Therefore, nuclear eigenstates have well-defined total isospin $T$ and isospin projection $T_z$. Consequences of this approximate isospin symmetry can be seen, for example, by comparing the energy levels of different nuclei with the same total number of nucleons. In addition, when modeling the interactions between two nucleons, we are mainly concerned with only two isospin states instead of dealing with four charge states. 

In this work, we carry out a detailed analysis of the implications of isospin
symmetry on two-body distribution functions. The latter encode nuclear correlation effects and are key quantities to model the interaction of atomic nuclei with electroweak probes. Two-body momentum distributions provide important input for the description of correlation effects in electron-scattering experiments \cite{Weiss:2018tbu,Andreoli:2021cxo}. In coordinate space, information regarding short-range correlations (SRCs) can be extracted from two-body densities \cite{Cruz-Torres2020}. As another example, neutrinoless double beta ($0\nu\beta\beta$) decay matrix elements can be directly calculated using appropriate two-body transition densities \cite{Cirigliano_2022_snowmass,Matteo2022,YAO2022103965,Simkovic2008,Simkovic2018,Cirigliano:2019vdj,Weiss2022_0nbb}. Finally, two-body current contributions in electroweak transitions can be expressed in terms of transition densities that exhibit a universal behavior at short distances~\cite{King:2020wmp}.  

In this work, we derive analytic relations between different diagonal and off-diagonal two-body distribution functions, including two-body densities, momentum distributions, and $0\nu\beta\beta$ decay transition densities, of a given nucleus or of different nuclei in the same isospin multiplet. These relations are completely general, as they do not depend on the specific nuclear Hamiltonian of choice, provided that isospin-breaking terms are small. Our relations enable sanity checks in ab-initio many-body calculations of diagonal and off-diagonal two-body density distributions. They are also relevant for experimental investigations of nuclear SRCs, i.e. nucleon pairs found in close proximity inside the nucleus. 

This paper is organized as follows. In Section~\ref{sec:expan} we present a general expansion of nuclear wave functions assuming isospin symmetry to be exact. In Section~ \ref{sec:relations}, we derive analytic relations between different diagonal and off-diagonal two-body distribution functions. Their implications to the study of nuclear SRCs are discussed in Section~\ref{sec:SRC}. Finally, in Section~\ref{sec:summary} we summarize and discuss the importance of our findings.  

\section{wave function expansion} 
\label{sec:expan}

We consider a nuclear wave function $\Psi$ of $A$ nucleons
with $T$ and $T_z$ isospin quantum numbers, neglecting the effect
of isospin-breaking terms. To analyze two-body quantities,
it is convenient to expand it using a complete set of orthonormal anti-symmetric two-body functions of the form
\be \label{eq:basis}
\varphi_m^{t,t_z}(\bs{r}_{ij}) \equiv \varphi_m^t(\bs{r}_{ij})\eta_{t,t_z},
\ee
where $t$ is the isospin of the pair, $t_z$ its projection,
$m$ denotes all the remaining quantum numbers,
and $\bs{r}_{ij}=\bs{r}_i-\bs{r}_j$ is the relative coordinate.
The radial and spin part is given by $\varphi_m^t(\bs{r})$,
while the isospin part is given by $\eta_{t,t_z}$. The latter is obtained from the coupling of two isospin-half nucleons and is normalized to 1, i.e. $\eta_{t,t_z}^\dagger \eta_{t,t_z} = 1$.

Expanding $\Psi$ using this basis we obtain
\begin{align} \label{eq:psi_expan}
& \Psi(\bs{r}_1,...,\bs{r}_A) \nonumber\\
& = \sum_{m} \varphi_m^{1}(\bs{r}_{12})
\sum_{t_z=-1}^1 \eta_{1,t_z} S_{m}^{1,t_z}(\bs{R}_{12},\bs{r}_3,...,\bs{r}_A)
\nonumber \\ &
+ \sum_{m} \varphi_m^{0}(\bs{r}_{12}) \eta_{0,0}
S_{m}^{0,0}(\bs{R}_{12},\bs{r}_3,...,\bs{r}_A),
\end{align}
where $\bs{r}_i$ are single-nucleon coordinates,
$\bs{R}_{ij}=(\bs{r}_i+\bs{r}_j)/2$ is the center-of-mass (CM) coordinate of the pair, and $S_{m}^{t,t_z}$ serve as the coefficients in this
expansion. Note that the superscripts $t,t_z$ and subscript $m$ in
$S_{m}^{t,t_z}$ are just indices and do not represent its quantum numbers. We stress that this is an exact expansion of $\Psi$ as the functions $\varphi_m^{t,t_z}(\bs{r}_{ij})$ form a complete basis. In particular, the anti-symmetry of $\Psi$ for the exchange of any two nucleons is respected in this expansion.

Since $\varphi_m^{t}$ are orthogonal, we conclude that
the functions 
\be \label{eq:t1_TTz_func}
\sum_{t_z=-1}^1
\eta_{1,t_z} S_{m}^{1,t_z}(\bs{R}_{12},\bs{r}_3,...,\bs{r}_A)
\ee
and
\be \label{eq:t0_TTz_func}
\eta_{0,0}
S_{m}^{0,0}(\bs{R}_{12},\bs{r}_3,...,\bs{r}_A)
\ee
must each separately have the same $T,T_z$ quantum numbers
as $\Psi$, for any $m$ (see Appendix \ref{sec:TTz} for more details).
In the second expression, $\eta_{0,0}$ is an isospin-zero function. Therefore, 
$S_{m}^{0,0}$ must be an isospin eigenstate with $T,T_z$ quantum numbers and we will thus use the notation $S_{m}^{0,0}(T,T_z)$.
Regarding the first expression, because it has $T,T_z$ quantum numbers,
it can be written as 
\begin{align}
&\sum_{t_z=-1}^1 
\eta_{1,t_z} S_{m}^{1,t_z}
=
\sum_{T^{A-2}} \left[ \tilde{S}_m(T^{A-2}) \otimes \eta_1 \right]^{T,T_z}
\nonumber \\ &=
\sum_{t_z} \eta_{1,t_z} \sum_{T^{A-2}} 
\langle T^{A-2} (T_z-t_z) 1 t_z | T T_z \rangle \tilde{S}_m(T^{A-2},T_z-t_z).
\end{align}
The sum over $T^{A-2}$ accounts for the different possible values
of the $A-2$ nucleon isospin, such that coupled to $t=1$ it can produce
the total isospin $T$. 
 $\tilde{S}_m(T^{A-2},T_z^{A-2})$ is a set of functions with well-defined
isospin quantum numbers of the $A-2$ nucleons. Their dependence
on $\bs{R}_{12},\bs{r}_3,...,\bs{r}_A$ is suppressed here.
$\bra T^{A-2} T_z^{A-2} t t_z | T T_z \ket$ are Clebsch-Gordan (CG) coefficients.
The last equation allows us to express $S_m^{1,t_z}$ using the 
$\tilde{S}_m(T^{A-2},T_z^{A-2})$ functions
\be \label{Eq:S_to_tilde_S}
S_m^{1,t_z}=\sum_{T^{A-2}} \langle T^{A-2} (T_z-t_z) 1 t_z | T T_z \rangle
 \tilde{S}_m(T^{A-2},T_z-t_z).
\ee
This relation, with the given set of $\tilde{S}_m(T^{A-2},T_z^{A-2})$ functions,
is applicable for all nuclei in the isospin multiplet of $\Psi$, i.e. for any value 
of $T_z$. This is the starting point for the relations derived in this paper.

\section{two-body distributions} \label{sec:relations}

We start by deriving relations involving the probability 
of finding two nucleons with a relative distance $r$ and isospin
quantum numbers $t,t_z$, i.e. the two-body density $\rho_{t,t_z}(r)$ defined as
\be \label{eq:dens_def}
\rho_{t,t_z}(r) = \frac{A(A-1)}{2} \frac{1}{4\pi r^2}
 \bra \Psi | \delta(r-r_{12}) \hat{\cal{P}}_{12}^{t,t_z} | \Psi \ket.
\ee
$\hat{\cal{P}}_{12}^{t,t_z}$ is a $t,t_z$ projection
for the combined isospin of particles $1$ and $2$. 
$\rho_{t,t_z}(r)$ obeys $\int dr 4\pi r^2 \rho_{t,t_z}(r) = N_{t,t_z}$,
where $N_{t,t_z}$ is the total number of $t,t_z$ pairs in the nucleus.

In general, we derive here two kinds of relations. The first kind
connects different $\rho_{t,t_z}(r)$ densities of a given nucleus, while the
second kind connects different $\rho_{t,t_z}(r)$ densities of different nuclei
that are part of the same isospin multiplet.

\subsection{Relations for a single nucleus}
\label{sec:single_nuc_rel}

Starting with the first kind of relations, we focus on the $t=1$
densities. 
Using Eqs. \eqref{eq:psi_expan} and \eqref{Eq:S_to_tilde_S}
 and the orthogonality of $\tilde{S}_m$ with respect to $T^{A-2}$, we obtain
\begin{align} \label{eq:rho_t1}
\rho_{1,t_z}&(r) = \frac{A(A-1)}{2} \frac{1}{4\pi r^2}
\sum_{T^{A-2}} |\langle T^{A-2} (T_z-t_z) 1 t_z | T T_z \rangle|^2
\nonumber \\ &\times
\sum_{m,m'} \bra \varphi_m^{1} | \delta(r-r_{12}) |\varphi_{m'}^{1} \ket 
 \bra \tilde{S}_m(T^{A-2}) | 
 \tilde{S}_{m'}(T^{A-2}) \ket.
\end{align}
Notice that the $\bra \tilde{S}_m | \tilde{S}_{m'} \ket$
matrix element is independent of the $T_z^{A-2}$ component,
due to the Wigner-Eckart theorem, and, therefore, it is suppressed above.
As a result, the only dependence on $t_z$ appears in the CG
coefficients. This observation allows us to derive relations between different
$t_z$ densities for certain values of $T$.

Starting with $T=0$ nuclei, the only possible 
contribution comes from $T^{A-2}=1$. In addition,
${|\langle 1 -t_z 1 t_z | 0 0 \rangle|^2=1/3}$ for all $t_z=\pm1,0$.
Therefore, the $t_z$ dependence disappears and we obtain
the expected relation for $T=0$ nuclei
\be \label{eq:T_zero}
\rho_{1,1}(r) = \rho_{1,0}(r) = \rho_{1,-1}(r)  \;\;\;\;\;\;\;   (T=0 \; \textrm{nuclei}),
\ee
i.e. the $pp$, $nn$, and $t=1$ $pn$ densities are equal.
The $pp$ and $nn$ densities, but not necessarily the $t=1$ $pn$ density $\rho_{1,0}$,
are expected to be identical for any state with $T_z=0$ (and any integer value of $T$).
This is simply obtained using this formulation based on the CG identity 
\be
|\bra T^{A-2}\; -t_z \;1\; t_z | T \;0 \ket|^2
=
|\bra T^{A-2}\; t_z \;1\; -t_z | T\; 0 \ket|^2,
\ee
relating the $t_z=1$ and $t_z=-1$ densities.

We move now to $T=1/2$ nuclei. In this case, both $T^{A-2}=1/2$ and $T^{A-2}=3/2$ contribute.
Looking at Eq. \eqref{eq:rho_t1}, the three $\rho_{1,t_z}(r)$ densities, 
for $t_z=\pm1,0$, are expressed using only two quantities 
of the form
\be \label{eq:A-2_contr}
\sum_{m,m'} \bra \varphi_m^{1} | \delta(r-r_{12}) |\varphi_{m'}^{1} \ket 
 \bra \tilde{S}_m(T^{A-2}) | 
 \tilde{S}_{m'}(T^{A-2}) \ket,
\ee
for $T^{A-2}=1/2,3/2$.
Therefore, there must be a linear relation between 
these three densities. A simple calculation, for both
$T_z=\pm 1/2$, leads to the result (see Appendix \ref{sec:Thalf_single_nucleus} for more details):
\be \label{eq:T_half}
2 \rho_{1,0}(r) = \rho_{1,1}(r)+\rho_{1,-1}(r)  \;\;\;\;\;\;\;  (T=\frac{1}{2} \; \textrm{nuclei}),
\ee
i.e. the $t=1$ $pn$ density is the average of the $pp$ and $nn$
densities for $T=1/2$ nuclei.
The same relation holds also if the densities are projected
to specific spin $s=0$ or $s=1$ of the pair.
This is a non-trivial result of isospin symmetry.
We can of course also integrate over the densities
and obtain expressions for the total number of $t=1$ and $t=0$ pairs.
They are consistent with the more general formulas
for the number of $t=1$ and $t=0$ pairs derived in Ref. \cite{Forest1996} 
for any value of $T$, see Eqs. (6.1) and (6.2).


For $T\geq 1$ there are three possible values of $T^{A-2}$.
There are also three $\rho_{1,t_z}(r)$ densities. 
Therefore, generally, no internal relation between the $\rho_{1,t_z}(r)$ densities
can be obtained based on isospin symmetry.
This is also a non-trivial statement regarding the implications of isospin
symmetry: the three $\rho_{1,t_z}(r)$ densities of a given nucleus are related to each other
only for $T=0$ and $T=1/2$ nuclei (or $T_z=0$ nuclei, for which the $pp$ and $nn$ densities are equal $\rho_{pp}=\rho_{nn}$).

We numerically test the above relations against variational Monte Carlo (VMC) calculations of light nuclei. The VMC method~\cite{Carlson:2014vla} approximates the solution of the nuclear Schr\"odinger equation with a trial wave function of the form
\begin{equation}
|\Psi_T\rangle = \Big(1 + \sum_{i<j<k}F_{ijk}\Big)\Big(\mathcal{S}\prod_{i<j}F_{ij}\Big) |\Psi_J\rangle\, 
\end{equation}
where $F_{ij}$ and $F_{ijk}$ are two- and three-body correlation operators, respectively. The symbol $\mathcal{S}$ indicates a symmetrized product over nucleon pairs since, in general, the spin-isospin dependent $F_{ij}$ do not commute. The strong $\alpha$-cluster  structure of light nuclei is explicitly accounted for by the antisymmetric Jastrow wave function $|\Psi_J\rangle$ that is constructed from a sum over independent-particle terms, each having four nucleons in an $\alpha$-like core and the remaining $(A-4)$ nucleons in p-shell orbitals~\cite{Pieper:2002ne}. The optimal set of variational parameters defining $F_{ij}$, $F_{ijk}$, and $|\Psi_J\rangle$ is found by minimizing the expectation value of the energy
\begin{equation}
E_T \equiv \frac{\langle \Psi_T | H | \Psi_T \rangle }{\langle \Psi_T | \Psi_T \rangle } \geq E_0, 
\end{equation}
where $E_0$ is the true ground-state energy of the system. Evaluating the above expectation value requires carrying out a multi-dimensional integration over the $3 A$ spatial coordinates of the nucleons. This is done in a stochastic fashion using the Metropolis-Hastings algorithm~\cite{Metropolis:1953,Hastings:1970}. 
 
\begin{figure}\begin{center}
\includegraphics[width=8.6 cm]{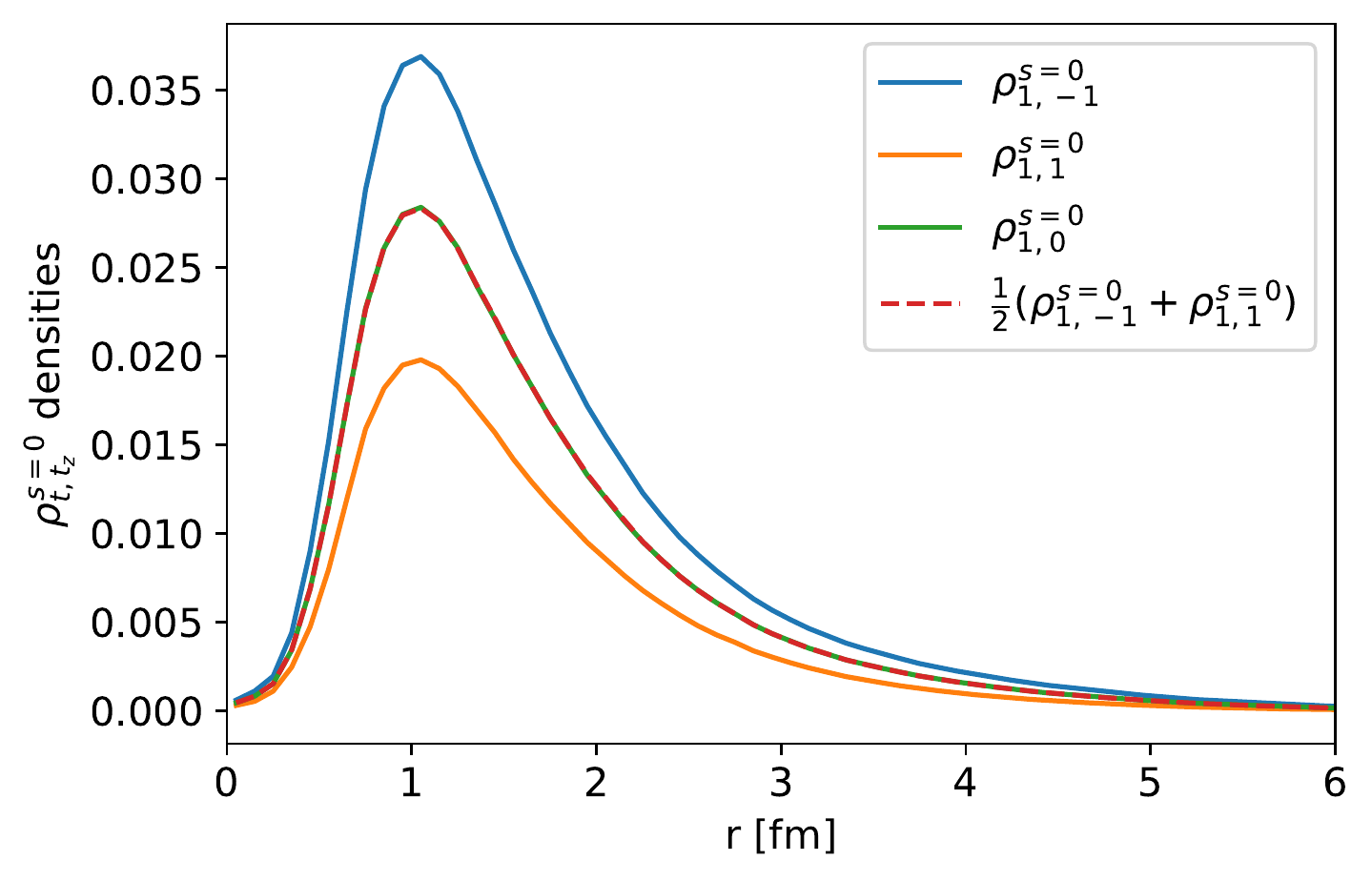}
\includegraphics[width=8.6 cm]{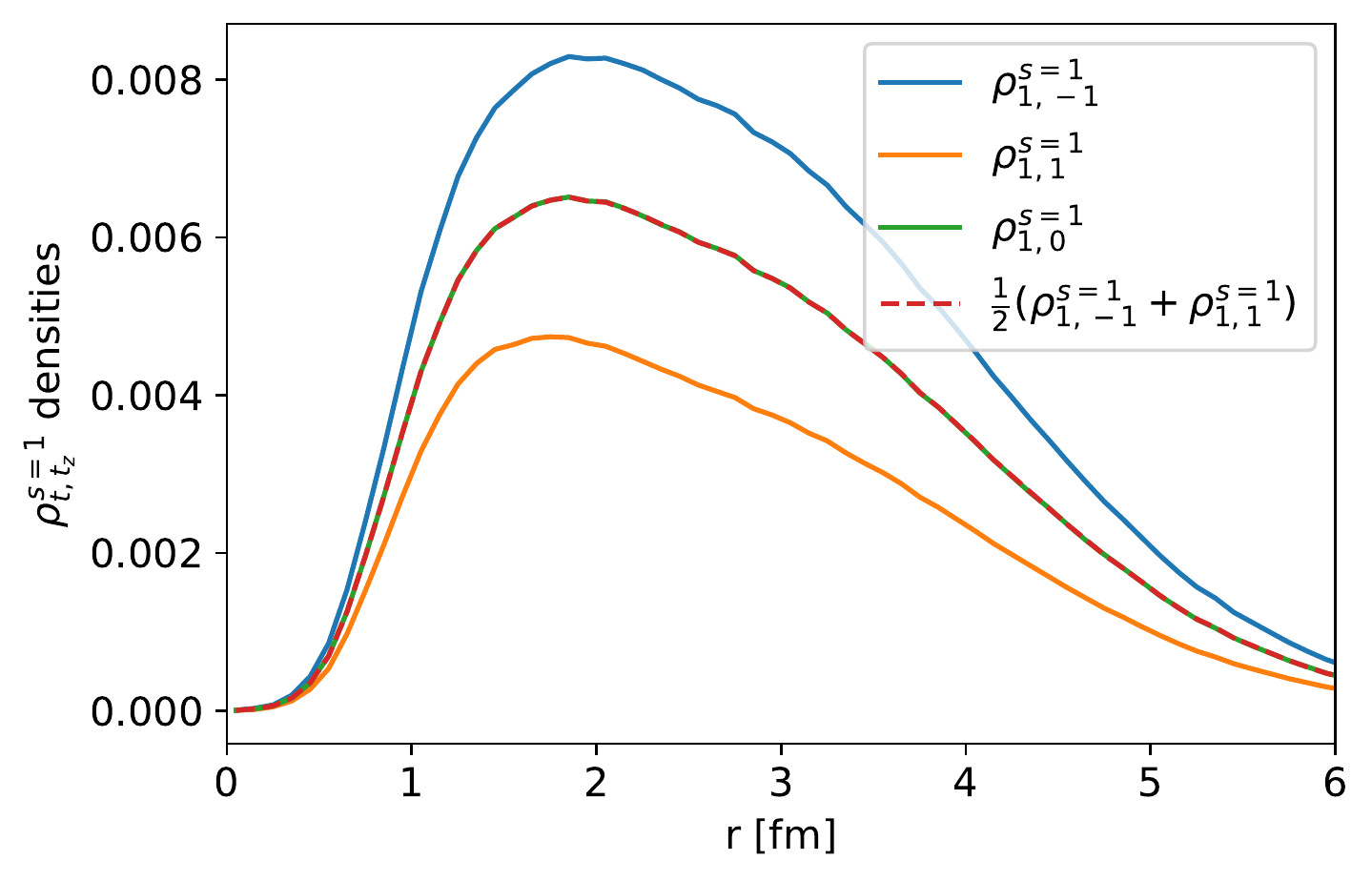}
\caption{\label{Li7_rho_NN} 
$t=1$ VMC two-body densities for $^7$Li using the AV18+UX potential,
with spin $s=0$ (top) and $s=1$ (bottom).
}
\end{center}\end{figure}

Calculations for $^7$Li $(T=1/2)$ are presented in Fig. \ref{Li7_rho_NN}. 
The $T=1/2$ relation, Eq. \eqref{eq:T_half}, is satisfied to a very good accuracy for both $s=0$
and $s=1$. 
Notice that the calculations are performed using the Argonne v18 (AV18) two-body interaction \cite{Wiringa:1994wb},
including the full electromagnetic interaction, and the Urbana X (UX) three-body interaction, which is a truncation of the Illinois-7 model \cite{Pieper2001}.
These models include small isospin-breaking terms, but explicit isospin-breaking operators are not included in the $F_{ij}$ or $F_{ijk}$.

\subsection{Relations for nuclei in the same isospin multiplet}
\label{sec:multiplet_rel}

We now discuss a second kind of relations, which connect densities of different nuclei in the same isospin multiplet.

We can start with the $t=0$ case, i.e. isospin-zero $pn$ pairs.
Based on Eqs. \eqref{eq:psi_expan} and \eqref{eq:dens_def}
we obtain an expression for the $t=0$ two-body density for a nucleus 
with $T,T_z$ quantum numbers
\begin{align}
\rho_{0,0}(r) &= \frac{A(A-1)}{2} \frac{1}{4\pi r^2}
\nonumber \\ &\times
\sum_{m,m'} \bra \varphi_m^{0} | \delta(r-r_{12}) |\varphi_{m'}^{0} \ket 
 \bra S_m^{0,0}(T,T_z) | S_{m'}^{0,0}(T,T_z) \ket.
\end{align}
The isospin quantum numbers of $S_m^{0,0}$, as identified
above,  are written here explicitly. Based on the Wigner-Eckart theorem the matrix element
$ \bra S_m^{0,0}(T,T_z) | S_{m'}^{0,0}(T,T_z) \ket$ does not depend on $T_z$.
Therefore, $\rho_{0,0}$ does not depend on $T_z$ and we conclude that the $t=0$
two-body density is identical for all nuclei in the same isospin multiplet, assuming
isospin symmetry. This remains true if the densities are projected to a specific spin $s$.

\begin{figure}\begin{center}
\includegraphics[width=8.6 cm]{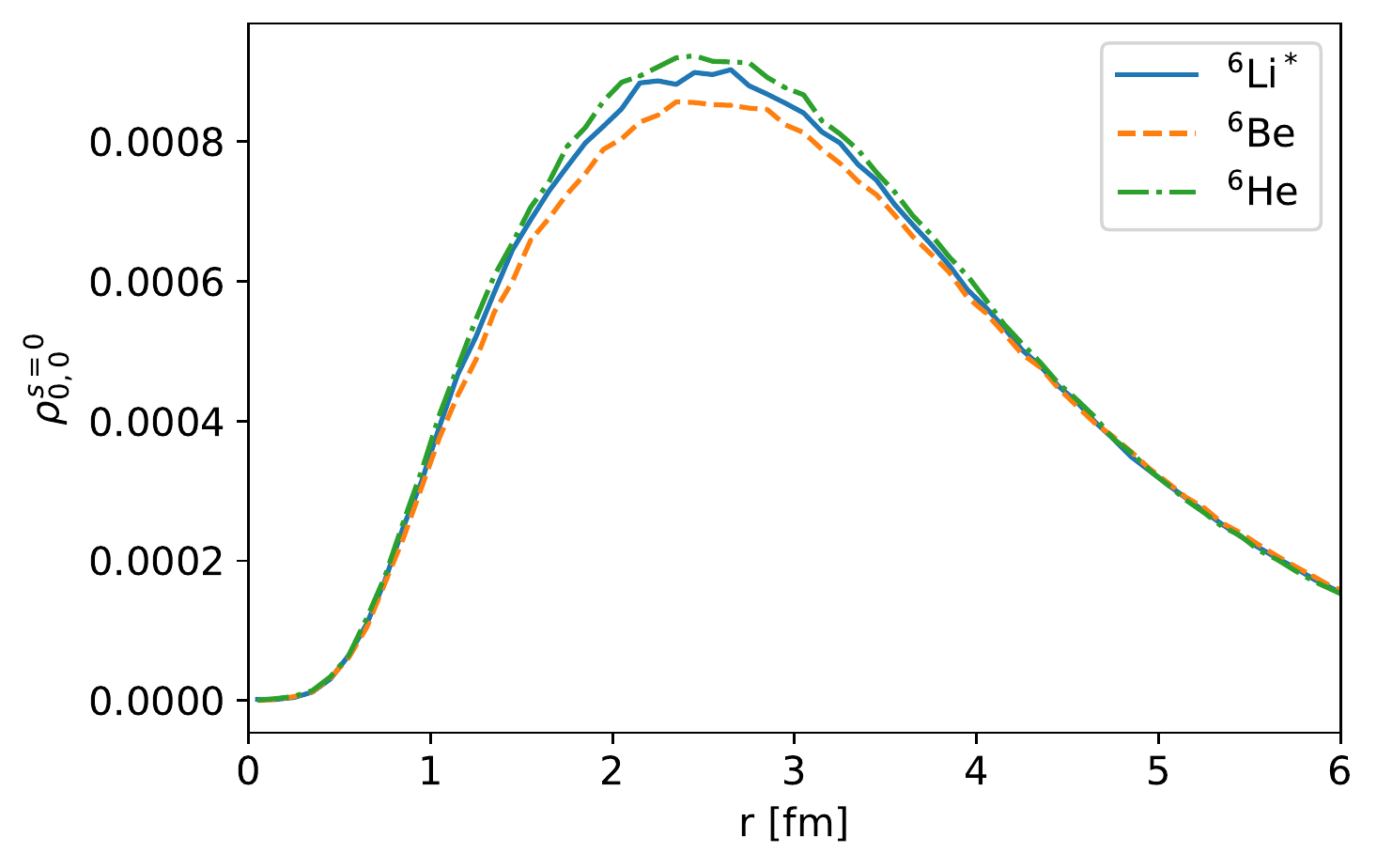}
\includegraphics[width=8.6 cm]{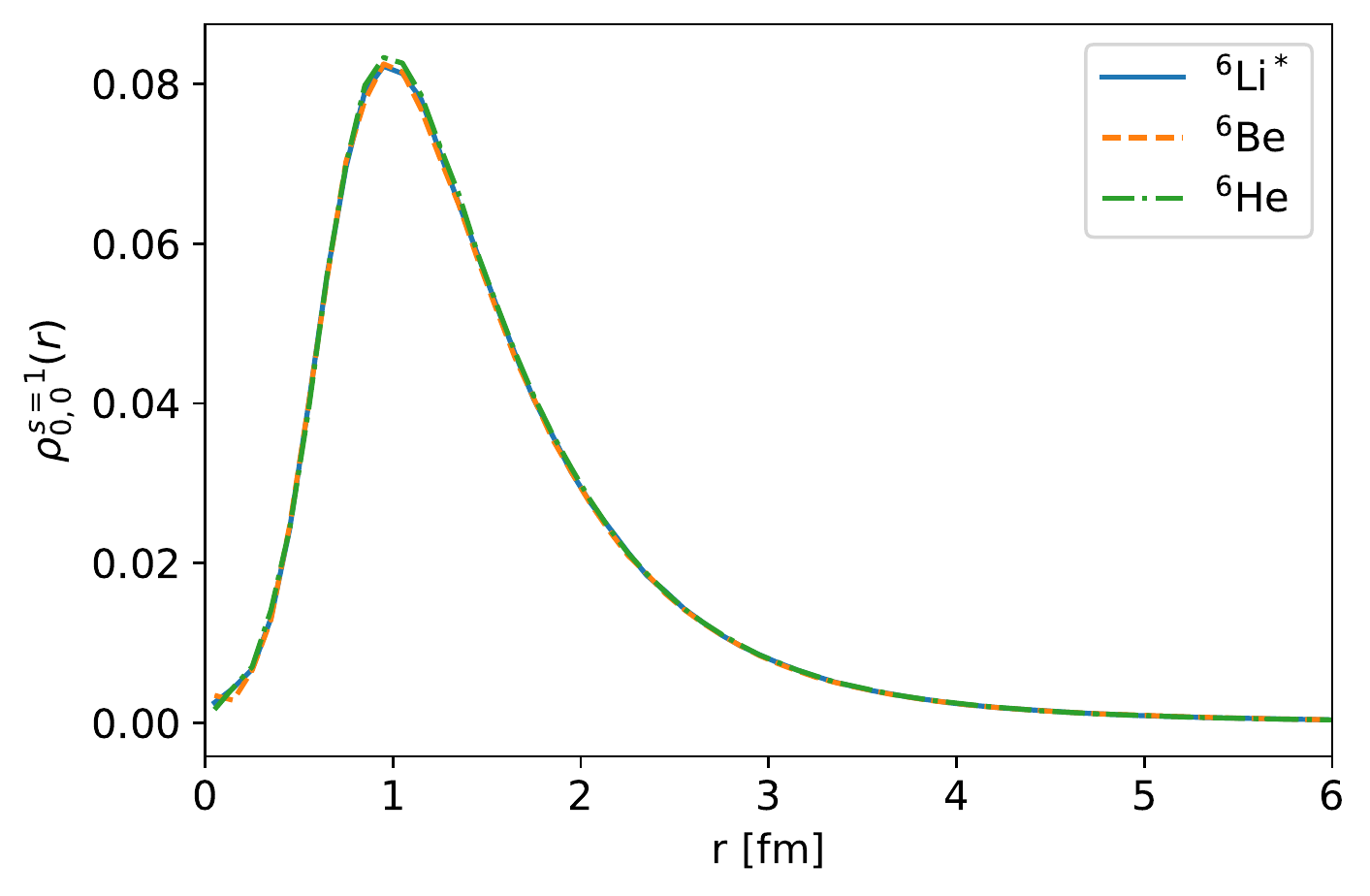}
\caption{\label{6He_6Lis_6Be_rho_NN_t0} 
$t=0$ VMC two-body densities for $^6$Be, $^6$Li*, and $^6$He
using the AV18+UX potential,
with spin $s=0$ (top) and $s=1$ (bottom).
}
\end{center}\end{figure}

As for the previous relations, we can verify the above relation through a comparison with VMC calculations.
Results for the $^6$He-$^6$Li*-$^6$Be 
$(J^\pi=0^+,T=1)$ multiplet, where $^6$Li* is an excited state, are displayed in Fig. \ref{6He_6Lis_6Be_rho_NN_t0}.
We can see that indeed the $t=0$ densities are equal for all nuclei
in this multiplet to a good accuracy,
with some small deviation for $s=0$, possibly due to the statistics of the
calculations or small isospin-breaking effects.
Notice that this $s=0$ density is significantly smaller than the $s=1$
density for $t=0$ pairs.

Relations for the $t=1$ densities can also be derived.
The general expression for these densities is given in Eq. 
\eqref{eq:rho_t1}, where there are at most three relevant values of $T^{A-2}$.
Therefore, the three $t=1$ densities in a given nucleus,
corresponding to three $t_z$ values, depend on no more than three quantities
of the form given in Eq. \eqref{eq:A-2_contr}.
Considering a specific $(T,T_z)$ nucleus, the three (or less) quantities
of Eq. \eqref{eq:A-2_contr} can be expressed using the three $\rho_{1,t_z}(r)$
densities. 
Notice that these quantities are identical for all nuclei in the same multiplet
and the only $T_z$ dependence appears in the 
CG coefficients in Eq. \eqref{eq:rho_t1}.
Therefore, the $\rho_{1,t_z}(r)$ densities of the remaining nuclei in the same multiplet can 
be directly expressed using the same densities of the $(T,T_z)$ nucleus, i.e.
knowing the three $\rho_{1,t_z}(r)$ densities of one nucleus allows us to obtain the
three  $\rho_{1,t_z}(r)$ densities of any nucleus in the same isospin multiplet.

Such relations exist for any value of $T$. We 
provide here explicit relations for a few cases but other cases
can be easily derived.
We use here the notation $\rho_{t,t_z}(T,T_z)$
for the $t,t_z$ density in a $(T,T_z)$ nucleus, suppressing the 
$r$ dependence.
Following the above arguments, for
$T=1/2$ nuclei we obtain the following relations:
\begin{align}
\rho_{1,0}\left(\frac{1}{2},\frac{1}{2}\right)&=\rho_{1,0}\left(\frac{1}{2},-\frac{1}{2}\right),
\\ \nonumber
\\
\rho_{1,\pm1}\left(\frac{1}{2},\frac{1}{2}\right)&=\rho_{1,\mp1}\left(\frac{1}{2},-\frac{1}{2}\right),
\end{align}
i.e. the $pn$ $t=1$ density is identical for the $T_z = \pm 1/2$ nuclei
while the $pp$ density in one nucleus is identical to the $nn$ density in the other.
These relations are very much expected as these two nuclei are mirror nuclei.
Using this formalism and CG identities, it can be shown that they
hold for any pair of mirror nuclei, i.e. 
\begin{align} 
\label{eq:mirror_pn}
\rho_{1,0}\left(T,T_z\right)&=\rho_{1,0}\left(T,-T_z\right),
\\ \nonumber
\\
\label{eq:mirror_pp_nn}
\rho_{1,\pm1}\left(T,T_z\right)&=\rho_{1,\mp1}\left(T,-T_z\right).
\end{align}

Following the same ideas, much less trivial relations
are obtained for the $T=1$ multiplet:
\begin{align} \label{eq:T=1_a}
\rho_{1,1}(1,0)=\rho_{1,-1}(1,0)=\rho_{1,0}(1,1)
\end{align}
\begin{align} \label{eq:T=1_b}
\rho_{1,0}(1,0)=\rho_{1,1}(1,1)-\rho_{1,0}(1,1)+\rho_{1,-1}(1,1).
\end{align}
In this case, the $t=1$ densities of the $(T=1,T_z=0)$ nucleus
are expressed using the $t=1$ densities of the $(T=1,T_z=1)$ nucleus.
Notice that the former is most likely an excited state, while the latter can be a ground state. This allows us to access information on excited states using ground-state quantities.
Specifically, the $pp$ and $nn$ densities of the $(T=1,T_z=0)$ nucleus are equal to the $pn$ $t=1$ density of the $(T=1,T_z=1)$ nucleus. The $pn$ $t=1$ density of the $(T=1,T_z=0)$ nucleus is equal to a specific linear combination of the $pp$, $nn$ and $pn$ $t=1$ densities of the $(T=1,T_z=1)$ nucleus.
Notice that the $(T=1,T_z=-1)$ nucleus is a mirror of the $(T=1,T_z=1)$ nucleus
and its densities can be obtained using Eqs. \eqref{eq:mirror_pn} and \eqref{eq:mirror_pp_nn}.
All these relations hold also if the densities are projected to a specific spin $s$.

For the $T=3/2$ multiplet we can similarly express the
excited-state $T_z=1/2$ densities using the $T_z=3/2$ densities:
\begin{align} \label{eq:T=3halfs_a}
\rho_{1,1}\left(\frac{3}{2},\frac{1}{2}\right) =
\frac{1}{3}\rho_{1,1}\left(\frac{3}{2},\frac{3}{2}\right)
+\frac{2}{3}\rho_{1,0}\left(\frac{3}{2},\frac{3}{2}\right)
\end{align}
\begin{align} \label{eq:T=3halfs_b}
\rho_{1,0}\left(\frac{3}{2},\frac{1}{2}\right) &=
\frac{2}{3} \rho_{1,1}\left(\frac{3}{2},\frac{3}{2}\right)
-\frac{1}{3} \rho_{1,0}\left(\frac{3}{2},\frac{3}{2}\right)
\nonumber \\ &
+\frac{2}{3} \rho_{1,-1}\left(\frac{3}{2},\frac{3}{2}\right)
\end{align}
\begin{align} \label{eq:T=3halfs_c}
\rho_{1,-1}\left(\frac{3}{2},\frac{1}{2}\right) =
\frac{2}{3} \rho_{1,0}\left(\frac{3}{2},\frac{3}{2}\right)
+\frac{1}{3} \rho_{1,-1}\left(\frac{3}{2},\frac{3}{2}\right).
\end{align}
Also here the relations for mirror nuclei allow us to obtain
the $T_z=-1/2$ and $T_z=-3/2$ densities,
and again all these relations hold also if the densities are projected to a specific spin $s$.

\begin{figure}\begin{center}
\includegraphics[width=8.6 cm]{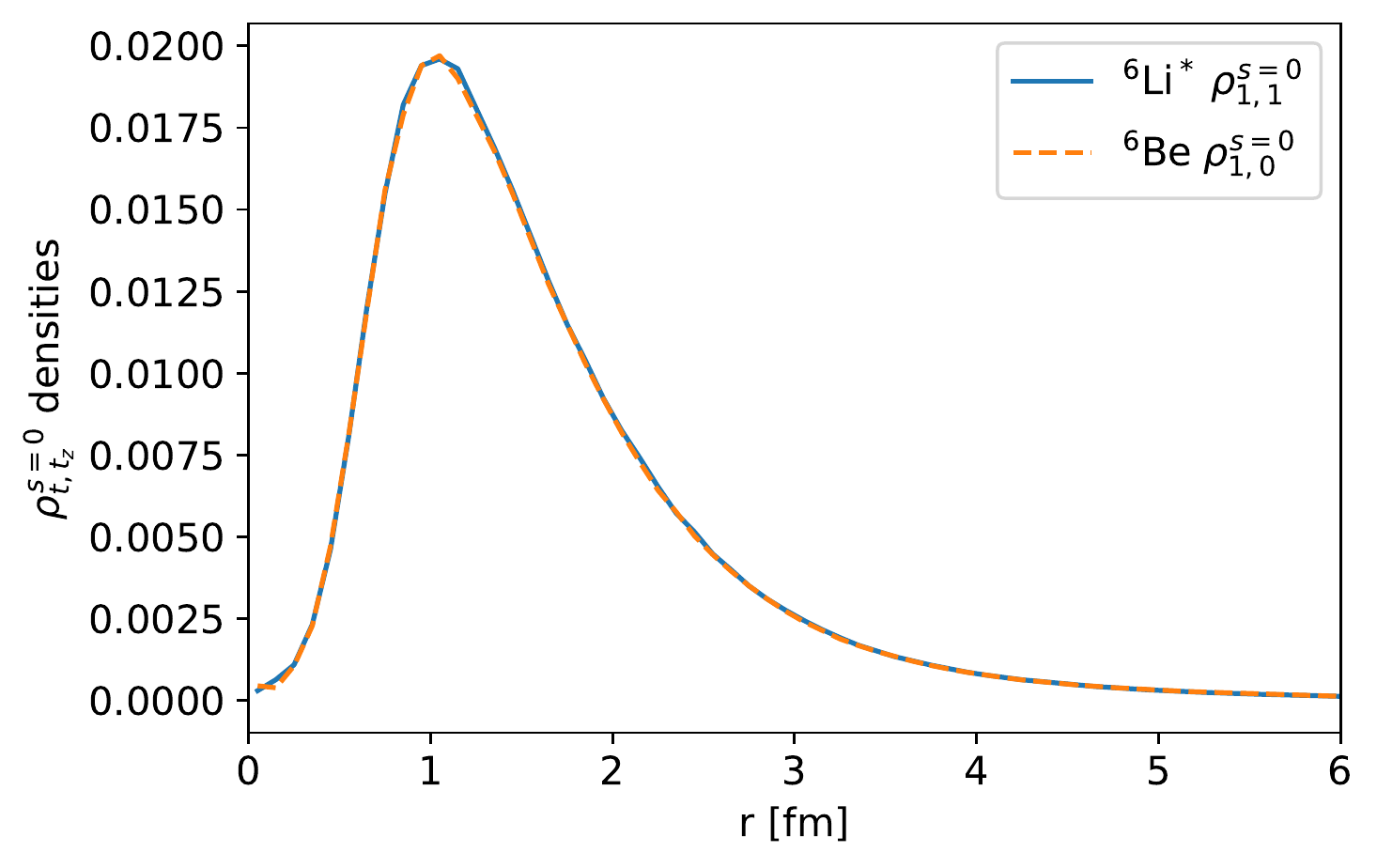}
\includegraphics[width=8.6 cm]{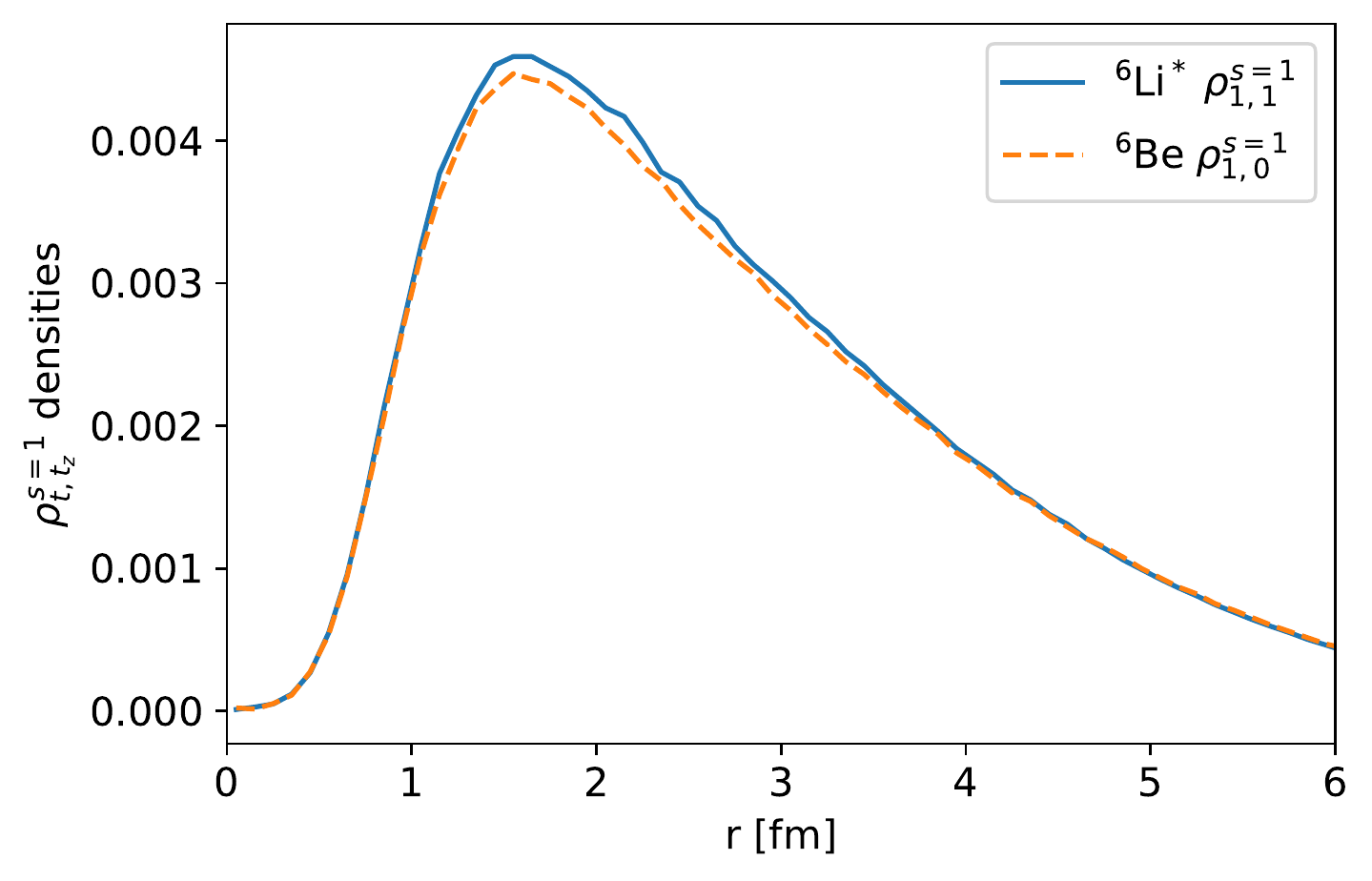}
\caption{\label{6He_6Lis_6Be_rho_NN_1st_relation} 
$t=1$ VMC two-body densities involved in Eq. \eqref{eq:T=1_a}
for $^6$Be and $^6$Li* using the AV18+UX potential,
with spin $s=0$ (top) and $s=1$ (bottom).
}
\end{center}\end{figure}

\begin{figure}\begin{center}
\includegraphics[width=8.6 cm]{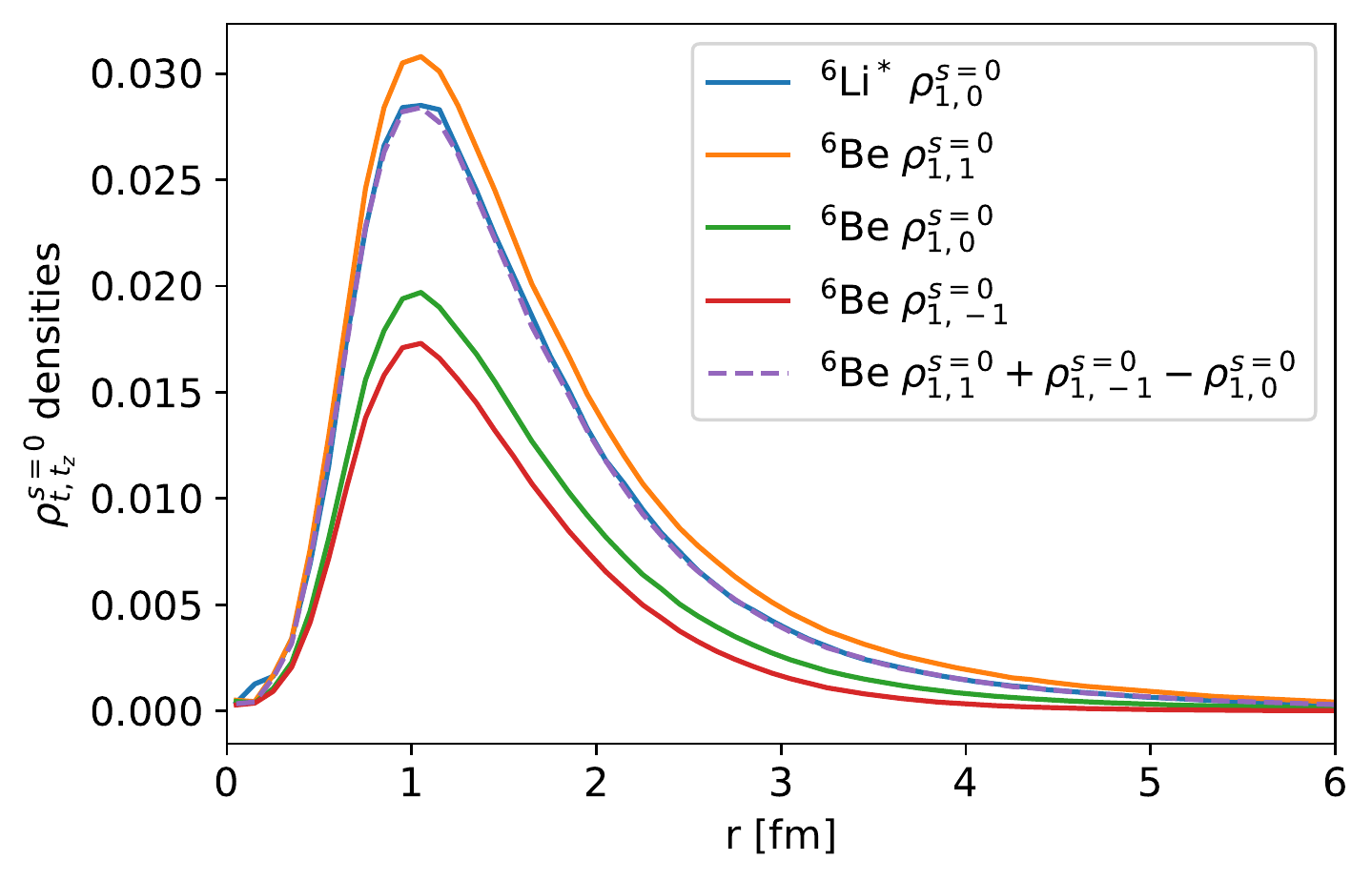}
\includegraphics[width=8.6 cm]{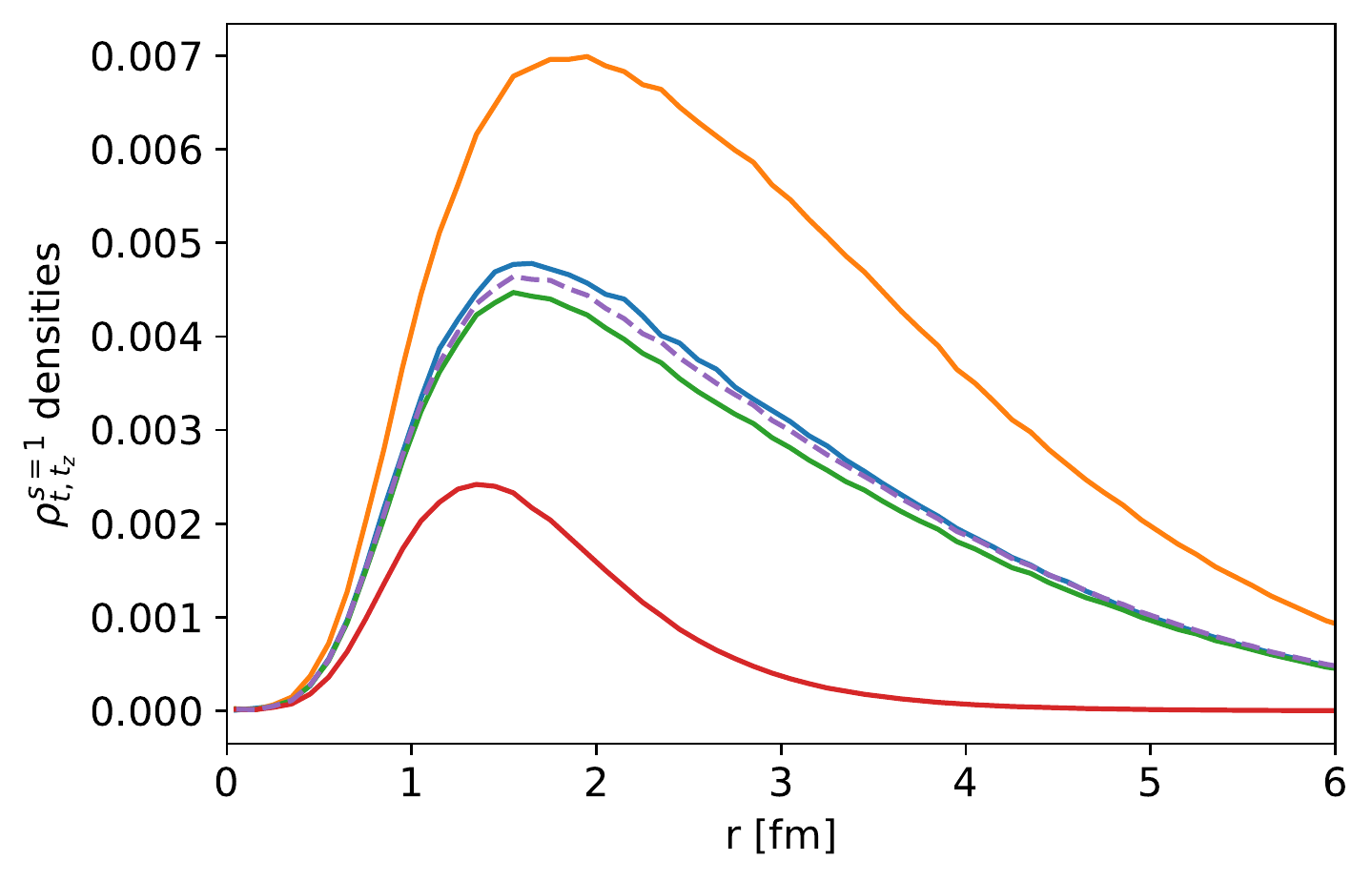}
\caption{\label{6He_6Lis_6Be_rho_NN_2nd_relation} 
$t=1$ VMC two-body densities involved in Eq. \eqref{eq:T=1_b}
for $^6$Be and $^6$Li* using the AV18+UX potential,
with spin $s=0$ (top) and $s=1$ (bottom). 
The legend of the bottom panel is identical to the top panel just with $s=1$.
}
\end{center}\end{figure}

We can provide a comparison to VMC calculations also for these relations.
Results for the $^6$He-$^6$Li*-$^6$Be 
$(J^\pi=0^+,T=1)$ multiplet are presented in Figs. \ref{6He_6Lis_6Be_rho_NN_1st_relation}
 and \ref{6He_6Lis_6Be_rho_NN_2nd_relation},
investigating the $T=1$ relations given in Eqs. \eqref{eq:T=1_a} 
and \eqref{eq:T=1_b}, respectively.
We can see that the relations are satisfied to a good accuracy,
with some small deviation for $s=1$, possibly due to isospin-breaking
effects or the statistics of the calculations.
Notice that this $s=1$ density is significantly smaller than the $s=0$
density for $t=1$ pairs.

\begin{figure}\begin{center}
\includegraphics[width=8.6 cm]{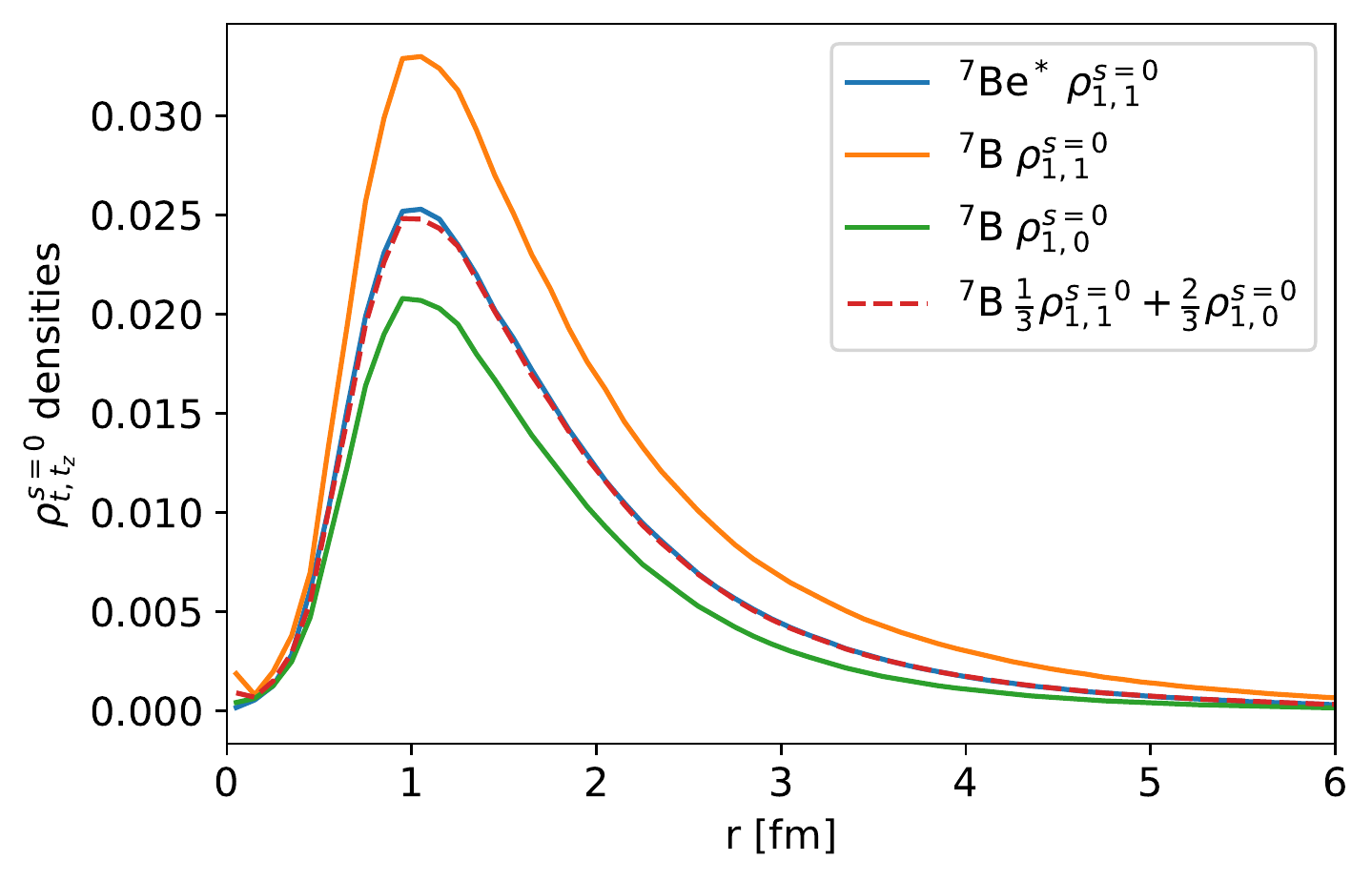}
\includegraphics[width=8.6 cm]{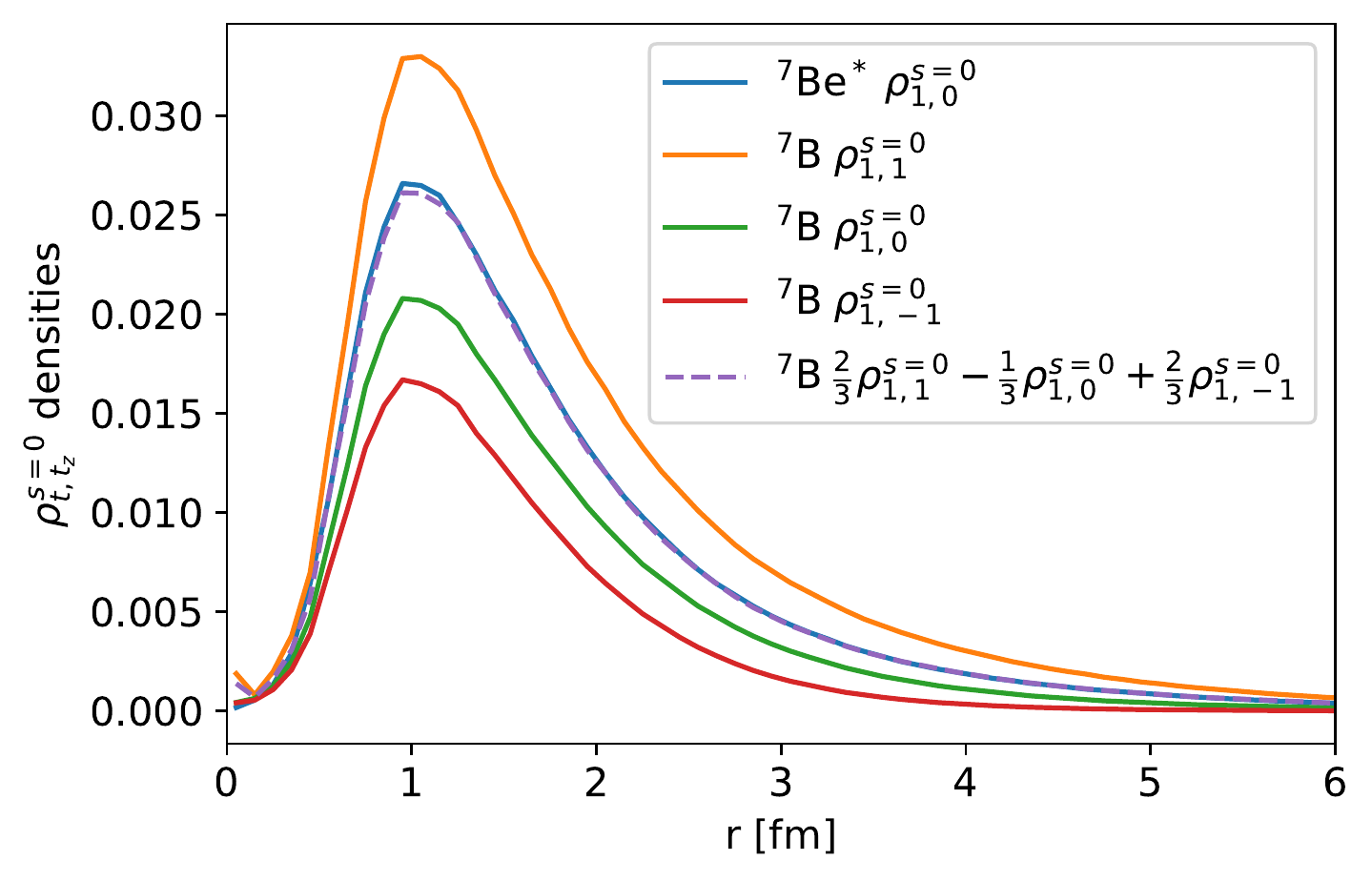}
\includegraphics[width=8.6 cm]{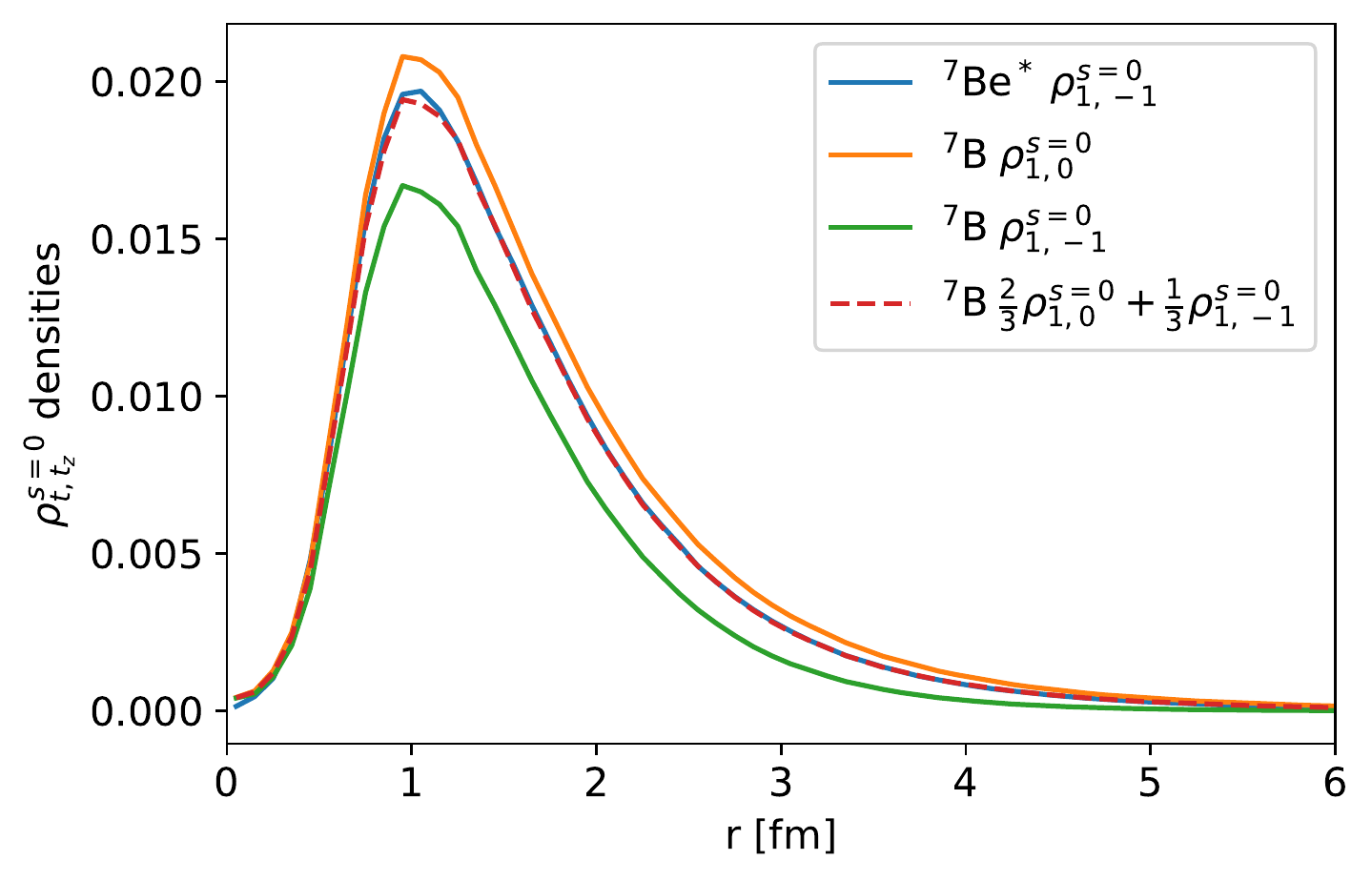}
\caption{\label{7Bes_7B_t1s0_relations} 
$t=1$ VMC two-body densities involved in Eqs. \eqref{eq:T=3halfs_a} (top), \eqref{eq:T=3halfs_b} (middle) and \eqref{eq:T=3halfs_c} (bottom)
for $^7$B and $^7$Be* using the AV18+UX potential,
with spin $s=0$.
}
\end{center}\end{figure}

Results for the $^7$He-$^7$Li*-$^7$Be*-$^7$B 
$(J^\pi=3/2^-,T=3/2)$ multiplet are shown in Fig.
\ref{7Bes_7B_t1s0_relations}. Once again we observe that VMC calculations obey the relations given in Eqs. \eqref{eq:T=3halfs_a}, \eqref{eq:T=3halfs_b} and \eqref{eq:T=3halfs_c} for spin $s=0$. The $s=1$ calculations (not shown here) are also in good agreement with some small deviations, similar to the $T=1$ case. 

Looking on the above $T=1/2$, $T=1$ and $T=3/2$ relations we can see that the sum over three $\rho_{1,t_z}$ densities is the same for all nuclei in the multiplet, i.e. the sum
$\sum_{t_z}\rho_{1,t_z}(T,T_z)$ is independent of $T_z$. Based on Eq. \eqref{eq:rho_t1} and the orthogonality relation of the CG coefficients we can conclude that this is a general property valid for all values of $T$.

We note that all the relations discussed in Secs. \ref{sec:single_nuc_rel} and \ref{sec:multiplet_rel} for the two-body densities hold equivalently for two-body momentum distributions,
describing the probability of finding two nucleons with a relative
momentum $k$ and isospin quantum number $t,t_z$.

\subsection{$0\nu\beta\beta$ decay}

Isospin-symmetry also has implications regarding
the transition densities of $0\nu\beta\beta$ decay
between mirror nuclei. 
$0\nu\beta\beta$ decay is a beyond-Standard-Model (BSM)
process in which two neutrons decay into two protons and two
electrons without any accompanying anti-neutrinos. If measured, it will have significant
implications to our understanding of the origin of the neutrino mass and of the matter dominance in the universe (see Refs. \cite{Cirigliano_2022_snowmass,Matteo2022,YAO2022103965} for recent reviews).
Experimentally, there is a potential to observe such a decay only in certain nuclei,
all of them involve an isospin change of $\Delta T = 2$ between the initial
and final nucleus. Nevertheless, theoretical studies of $0\nu\beta\beta$ involve also decay between mirror nuclei ($\Delta T = 0$), e.g. for the purpose of testing different models \cite{Pastore:2017ofx,Wang:2019hjy,Yao:2020olm,Novario:2020dmr}.

Extracting quantitative BSM information from $0\nu\beta\beta$ decay measurement
requires the calculations of nuclear matrix elements (NMEs).
These NMEs can be calculated by integrating over transition densities
(multiplied by the appropriate neutrino potential and other factors),
that can be separated into Fermi (F), Gamow-Teller (GT), and tensor components. In this work, we focus on the first two transitions, which are defined as (see e.g. \cite{Simkovic2008,Simkovic2018,Cirigliano:2019vdj,Weiss2022_0nbb})
\be
\rho_F(r) = \frac{A(A-1)}{2}\frac{1}{4\pi r^2}
\bra \Psi_f | \delta(r-r_{12}) \tau_1^+ \tau_2^+ | \Psi_i \ket
\ee
\be
\rho_{GT}(r) = \frac{A(A-1)}{2}\frac{1}{4\pi r^2}
\bra \Psi_f | \delta(r-r_{12}) \bs{\sigma}_1\cdot\bs{\sigma}_2 \tau_1^+ \tau_2^+ | \Psi_i \ket.
\ee
Here, $\Psi_i$ and $\Psi_f$ are the initial and final nuclei,
and $\bs{\sigma}_a$ and $\bs{\tau}_a$ are the nucleon spin and
isospin operators of nucleon $a$, respectively.

\begin{figure}\begin{center}
\includegraphics[width=8.6 cm]{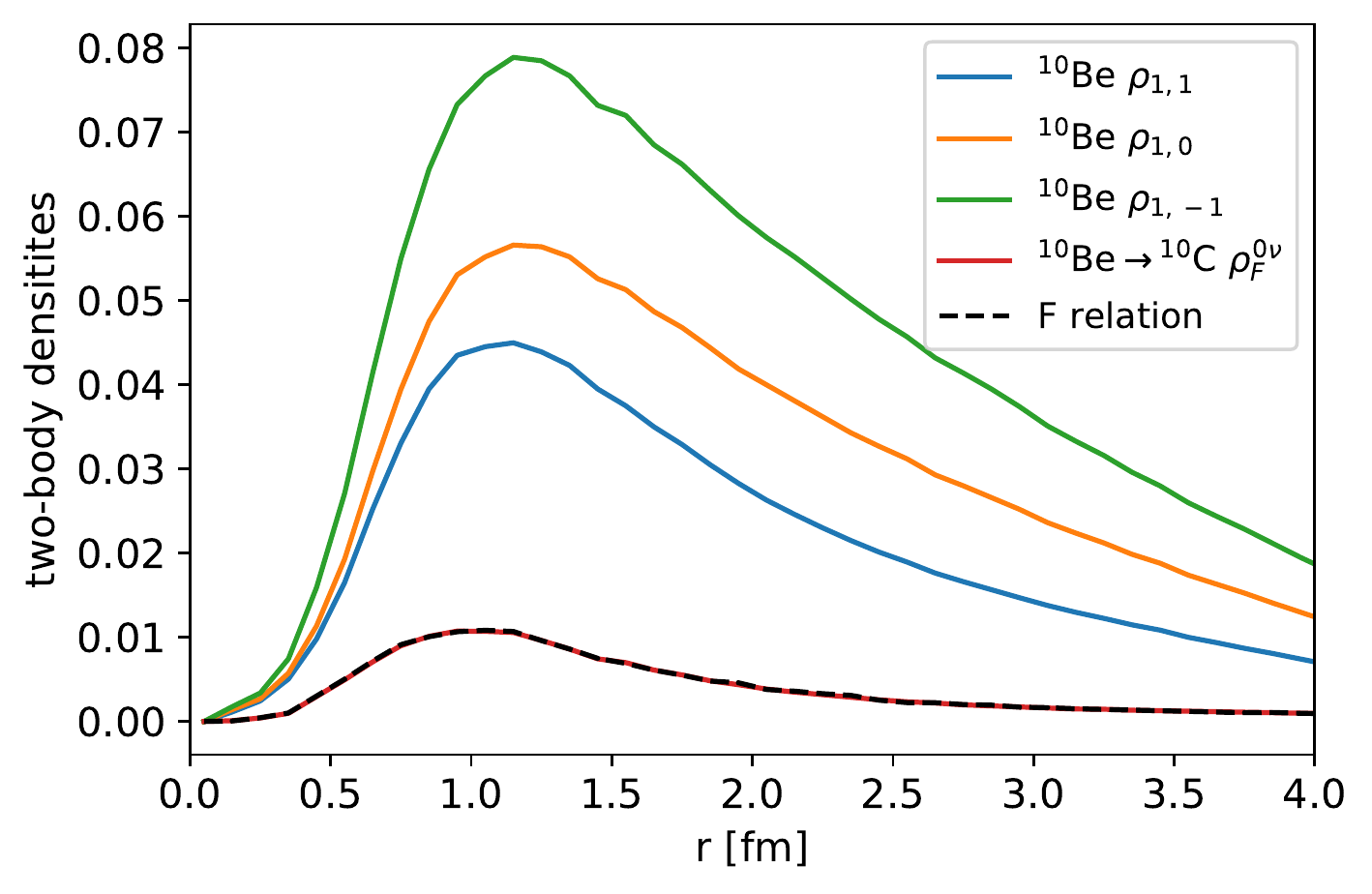}
\includegraphics[width=8.6 cm]{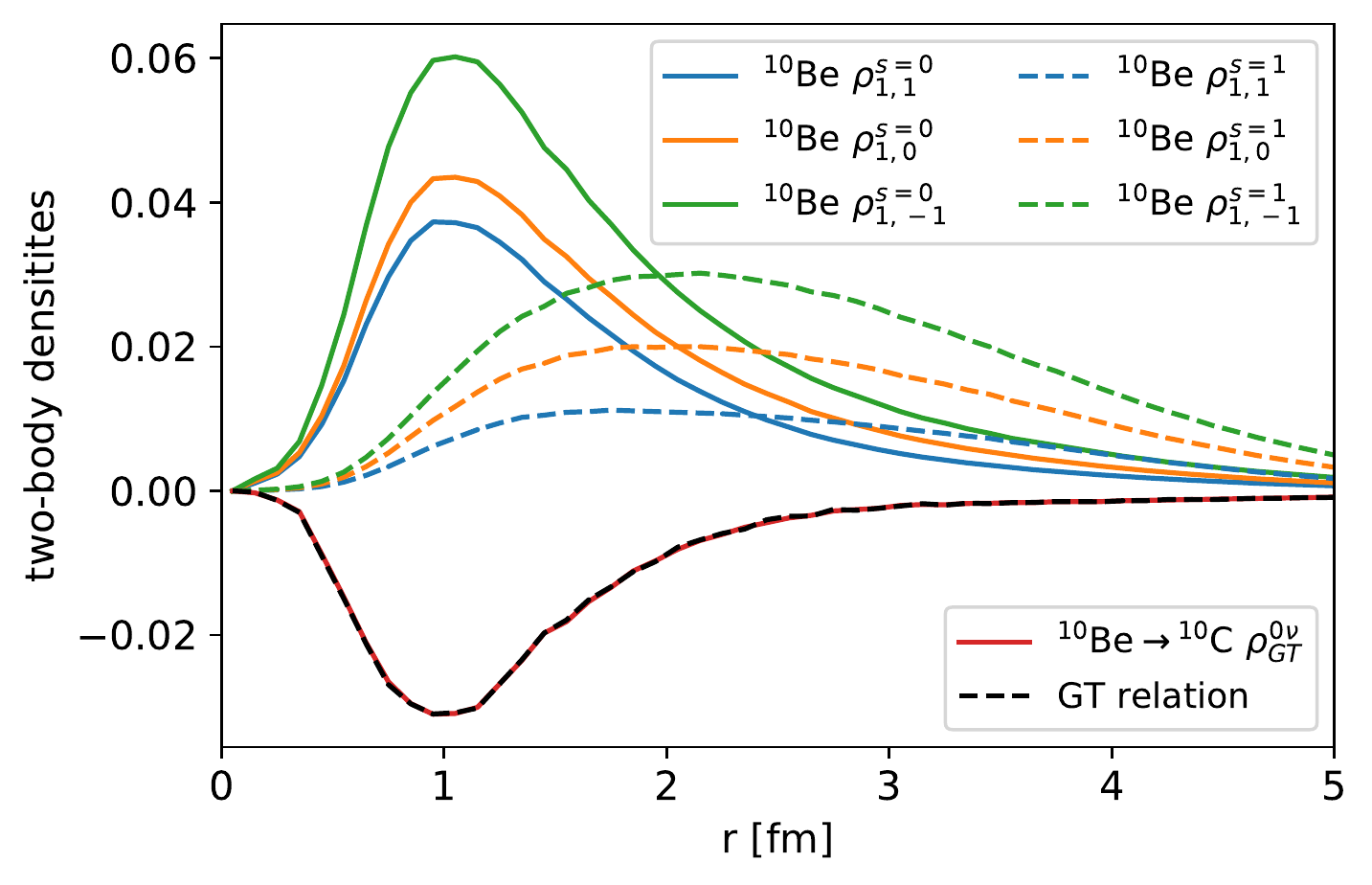}
\caption{\label{Be10_0nu2b} 
(top) $\rho_{1,t_z}$ two-body densities for $^{10}$Be and the 
$^{10}$Be$\rightarrow^{10}$C $0\nu\beta\beta$ Fermi transition density using the VMC method and the AV18+UX potential. The dashed line labeled as "F relation" shows the linear combination of $\rho_{1,t_z}$ for $^{10}$Be given in Eq. \eqref{eq:F_relation}.
(bottom) The same, but for the spin-projected $\rho_{1,t_z}^{s}$ two-body densities and the GT transition. The label "GT relation" refers to the linear combination of $\rho_{1,t_z}^{s}$ given in Eq. \eqref{eq:GT_relation}.
}
\end{center}\end{figure}

We consider here a decay between two ground-state
mirror nuclei. Therefore, these are necessarily $T=1$ nuclei,
where the initial and final states have isospin projection $T_z(i)=-1$
and $T_z(f)=1$, respectively.
The expansion given in Eq. \eqref{eq:psi_expan}
and the relation given in Eq. \eqref{Eq:S_to_tilde_S}
can be now utilized. For the Fermi transition we obtain
\begin{align} \label{eq:rho_F_expan}
\rho_F(r) &= \frac{A(A-1)}{8\pi r^2}
\sum_{T^{A-2}}  \langle T^{A-2} 0 1 1 | 1 1 \rangle^*
\langle T^{A-2} 0 1 -1 | 1 -1 \rangle
\nonumber \\ &\times
\sum_{m,m'} \bra \varphi_m^{1} | \delta(r-r_{12})| \varphi_{m'}^{1} \ket
\bra   \tilde{S}_m(T^{A-2}) |  \tilde{S}_{m^\prime}(T^{A-2})  \ket.
\end{align}
Notice that only $nn$ pairs in the initial state and $pp$ pairs
in the final state contribute.
There are three values of $T^{A-2}$ that contribute here
$(0,1, 2)$. In addition, the expression on the second line of
Eq. \eqref{eq:rho_F_expan} is identical to the expression appearing
in the two-body densities, Eq. \eqref{eq:rho_t1}.
As discussed above for the two-body densities, these expressions
for the three values of $T^{A-2}$ are directly related to the three
densities $\rho_{1,t_z}$ (of the initial state, for example).
Therefore, we obtain a relation between the Fermi $0\nu\beta\beta$
transition density of mirror nuclei and the two-body densities
\be \label{eq:F_relation}
\rho_F(r) = \rho_{1,-1}(r) + \rho_{1,1}(r) -2\rho_{1,0}(r), 
\ee 
where on the right-hand side we have here the densities of the initial-state nucleus.
A similar relation for the GT transition can also be derived,
involving spin-projected two-body densities $\rho_{t,t_z}^s$ to account for the effect of the $\sigma_a\cdot\sigma_b$ operator.
We obtain
\begin{align} \label{eq:GT_relation}
 \rho_{GT}(r)=
&-3 \left[\rho_{1,-1}^{s=0}(r) + \rho_{1,1}^{s=0}(r) - 2\rho_{1,0}^{s=0}(r)  \right]
\nonumber \\ &+
 \left[\rho_{1,-1}^{s=1}(r) + \rho_{1,1}^{s=1}(r) - 2\rho_{1,0}^{s=1}(r)  \right].
\end{align}
We can compare these relations to VMC calculations. Results of the $^{10}$Be$\rightarrow^{10}$C $\Delta T = 0$ transition are presented in Fig. \ref{Be10_0nu2b}. A remarkably good agreement is observed.

\section{short-range correlations} \label{sec:SRC}

In this section, we consider the implications of 
isospin symmetry on nuclear SRCs. 
The study of nuclear SRCs is focused on describing the properties
of nucleons that are found close to each other inside the nucleus,
and their impact on different nuclear quantities. Currently, there is 
a comprehensive understanding of the nature of two-nucleon SRCs,
i.e. pairs of nucleons at close proximity or with large relative momentum
and small CM momentum. Such pairs are predominantly back-to-back $pn$ pairs, due to the effect of the tensor force, and are the origin for most of high-momentum nucleons in the nucleus.
This picture is corroborated by both experimental measurements,
mainly high-energy quasi-elastic electron- and proton-scattering measurements, 
and theoretical studies, including ab-initio calculations of two-body densities
and momentum distributions --- see, e.g., Refs. \cite{Atti:2015eda,Hen:2016kwk,Arrington:2022sov,Wiringa:2013ala} and references therein.

The generalized contact formalism (GCF) is an effective theory
for describing SRCs. It is based on the asymptotic factorization of the nuclear
wave function, when two particles are close to each other, to a part describing
the correlated pair and another one describing the spectator $A-2$ nucleons \cite{Weiss:2015mba}:
\be
\Psi \xrightarrow[r_{ij}\rightarrow 0]{}
\sum_\alpha \varphi^\alpha(\bs{r}_{ij}) A^\alpha(\bs{R}_{ij},\{\bs{r}_k\}_{k\neq i,j}).
\ee
In the above equation, $\alpha$ denotes the quantum numbers of the pair, i.e. parity $\pi_\alpha$, spin $s_\alpha$, total angular momentum $j_\alpha$ and projection $j_{\alpha z}$, and total isospin $t_\alpha$
and projection $t_{\alpha z}$. The universal function 
$\varphi^\alpha(\bs{r}_{ij})$ is Hamiltonian dependent but nucleus independent. It describes the pair dynamics and is defined as the solution of the zero-energy two-body Schr\"odinger equation.
This factorization of the wave function was tested against ab-initio calculations \cite{Weiss:2015mba,Weiss:2016obx,Cruz-Torres2020},
and allows us to quantitatively study the properties of SRCs and their impact on different quantities such
as momentum distributions and two-body densities \cite{Weiss:2015mba,Weiss:2016obx,Cruz-Torres2020}, electron-scattering cross sections \cite{Weiss:2018tbu,schmidt20,Pybus:2020itv,Duer:2018sxh,weiss2020inclusive,CLAS:2020rue,Patsyuk:2021fju}, 
$0\nu\beta\beta$ matrix elements \cite{Weiss2022_0nbb}, the Coulomb sum rule \cite{Weiss:2016bxw} and photo-absorption cross sections \cite{Weiss:2014gua,Weiss:2015pjw}.
The GCF is also used in the planning of new experiments, e.g. involving photon-induced reactions \cite{GlueX:2020dvv} and for measurements in the future Electron Ion Collider \cite{Hauenstein:2021zql}.

The nuclear {\it contacts} for a nucleus with $A$ nucleons are defined as \cite{Weiss:2015mba}
\be
C^{\alpha\beta}=\frac{A(A-1)}{2}\langle A^\alpha | A^\beta \rangle.
\label{eq:contact}
\ee
The factor $A(A-1)/2$ appears in place of the number of proton-proton, neutron-proton or neutron-neutron pairs present in most previous publications because here the wave function is anti-symmetric under permutation of any two nucleons.
The diagonal contacts $C^{\alpha\alpha}$ are proportional to the number of correlated pairs in the nucleus with quantum numbers $\alpha$.

Among all possible channels, accounted for by the sum over $\alpha$, it was identified that the two channels that include an s-wave component are the most significant \cite{Weiss:2016obx,Cruz-Torres2020}.
These channels are the so-called isospin-zero deuteron channel, defined by positive parity, $s_\alpha=1$, $j_\alpha=1$, $t_\alpha=0$, and the isospin-one channel defined by positive parity,  $s_\alpha=0$, $j_\alpha=0$, $t_\alpha=1$. Focusing on the $t=1$ contribution, the factorized wave function can be written as
\be
\Psi \xrightarrow[r_{ij}\rightarrow 0]{}
\varphi^{t=1}(\bs{r}_{ij}) \sum_{t_z=-1}^1 
\eta_{1,t_z} A^{t=1}_{t_z}(\bs{R}_{ij},\{\bs{r}_k\}_{k\neq i,j}).
\ee
The index $t=1$ refers here specifically to the above-mentioned
leading isospin-one channel. The two-body function is independent
of $t_z$ assuming isospin symmetry, and the isospin part $\eta_{1,t_z}$
is written explicitly. 
The corresponding contact terms are given by
\be \label{eq:t1_contacts}
C^{t=1}_{t_z} = \frac{A(A-1)}{2}\langle A^{t=1}_{t_z} | A^{t=1}_{t_z} \rangle.
\ee
Therefore, the $t=1$ two-body density obey the following asymptotic
expression
\be
\rho_{t=1,t_z}(r) \xrightarrow[r_{ij}\rightarrow 0]{}
 C^{t=1}_{t_z} \frac{1}{4\pi r^2}
 \bra \varphi^{t=1} | \delta(r-r_{12}) | \varphi^{t=1} \ket.
\ee
Based on this expression, some of the results of the previous section
have direct implication on the contact values.
For example, Eqs. \eqref{eq:T_zero} and \eqref{eq:T_half}, which connect
the different $t=1$ densities for $T=0$ and $T=1/2$ nuclei, respectively,
imply similar relations for the nuclear contacts
\be 
C^{t=1}_{t_z=1} = C^{t=1}_{t_z=0} = C^{t=1}_{t_z=-1}  \;\;\;\;\;\;\;   (T=0 \; \textrm{nuclei}),
\ee
\be 
2  C^{t=1}_{t_z=0} = C^{t=1}_{t_z=1}+C^{t=1}_{t_z=-1}  \;\;\;\;\;\;\;  (T=\frac{1}{2} \; \textrm{nuclei}).
\ee
The $T=0$ relation was realized and utilized before \cite{Weiss:2016obx}, but the $T=1/2$ relation was not, and can be helpful in extracting more accurate contact values.
Knowing the contact values is crucial for analyzing the effect of SRCs on different observables.
This relation can be important for extracting information from inclusive and exclusive electron-scattering data, where disentangling the contribution of different channels can be difficult. In the inclusive case, the combined contribution from all pairs is measured. In exclusive measurements, where the knocked-out pair is detected, it is possible in principle to measure the $pp$, $nn$ and $pn$ contributions, but the above relation
can allow disentangling the isospin-one and isospin-zero contributions for $T=1/2$ nuclei.
Additional relations between nuclear contacts arise similarly from the 
relations connecting the densities of different nuclei in the same isospin multiplet,
Eqs. \eqref{eq:mirror_pn}-\eqref{eq:T=3halfs_c} 
(and their generalization to $T>3/2$ multiplets). These relations allow us to extract the $t=1$
nuclear contacts of all nuclei in an isospin multiplet if they are known for one nucleus in this multiplet.
The $t=0$ contact is the same for all nuclei in the multiplet. 
Therefore, nuclear contact values of excited states can be obtained using nuclear contact values of ground-state
nuclei in the same multiplet.

$0\nu\beta\beta$ transitions were recently analyzed using the GCF, introducing new contact parameters $C(f,i)$, describing the transition from the initial ($i$) to final ($f$) nucleus \cite{Weiss2022_0nbb}. The relations between the two-body densities and the $0\nu\beta\beta$ transition densities, given in Eqs. \eqref{eq:F_relation} and \eqref{eq:GT_relation}, lead to a direct relation between the $C_{t_z}^1$ contacts and the $C(f,i)$ $0\nu\beta\beta$ contacts for $\Delta T = 0$ transitions
\be
C(f,i) = C^{t=1}_{t_z=-1}(i)+C^{t=1}_{t_z=1}(i)-2C^{t=1}_{t_z=0}(i),
\ee
where on the right-hand side we have the contacts of the initial nucleus. Since $C_{t_z}^1$ contact values can be extracted from electron-scattering experiments \cite{schmidt20}, this relation provides a connection between such electron-scattering reactions and $\Delta T = 0$ $0\nu\beta\beta$ transitions.

On top of these direct consequences of the general two-body density
relations, isospin symmetry has an additional implication for the description of SRCs: it allows us to extract information regarding the isospin structure of the spectator $A-2$ nucleon system.
It can be seen most simply if we consider a $pn$ $t=0$ SRC pair. In this case, the
$A-2$ nucleon subsystem must have the same $T,T_z$ quantum numbers of the full wave function
$\Psi$. 
Considering a $t=1$ SRC pair, the analysis becomes non trivial.
For $T=0$ nuclei, the $A-2$ nucleon part must have $T^{A-2}=1$, but for
$T=1/2$ nuclei there are two possible values $T^{A-2}=1/2,3/2$ and for 
$T\geq 1$ there are three possible values.

Similar to the analysis in Sec. \ref{sec:expan}, based on isospin symmetry
we can write
\be
A^{t=1}_{t_z}=\sum_{T^{A-2}} \langle T^{A-2} (T_z-t_z) 1 t_z | T T_z \rangle \tilde{A}^{t=1}(T^{A-2},T_z-t_z),
\ee
for a nucleus with $T,T_z$ quantum numbers.
$\tilde{A}^{t=1}(T^{A-2},T_z-t_z)$ are functions with well-defined isospin quantum number
of the $A-2$ nucleon system.
Then, using Eq. \eqref{eq:t1_contacts}, the $t=1$ contacts are given by,
\begin{align} \label{eq:C_t1_A_tilde}
C_{t_z}^{t=1}
&=
\frac{A(A-1)}{2} 
\sum_{T^{A-2}} 
|\langle T^{A-2} (T_z-t_z) 1 t_z | T T_z \rangle |^2
\nonumber \\ &\times
\langle \tilde{A}^{t=1}(T^{A-2})| \tilde{A}^{t=1}(T^{A-2}) \rangle.
\end{align}
It has to be noted that the $\bra \tilde{A}^{t=1}| \tilde{A}^{t=1} \ket$
matrix element is independent of the $T_z^{A-2}$ component,
due to the Wigner-Eckart theorem. 
We see that the three $t=1$ contacts of a given nucleus depend on the same
elements $\langle \tilde{A}^{t=1}(T^{A-2})| \tilde{A}^{t=1}(T^{A-2}) \rangle$.
There are no more than three such elements, as discussed above.
Therefore, the relation can be reversed and we can express 
$\langle \tilde{A}(T^{A-2})| \tilde{A}(T^{A-2}) \rangle$ using the three 
$t=1$ contacts. 

\begin{figure}\begin{center}
\includegraphics[width=8.6 cm]{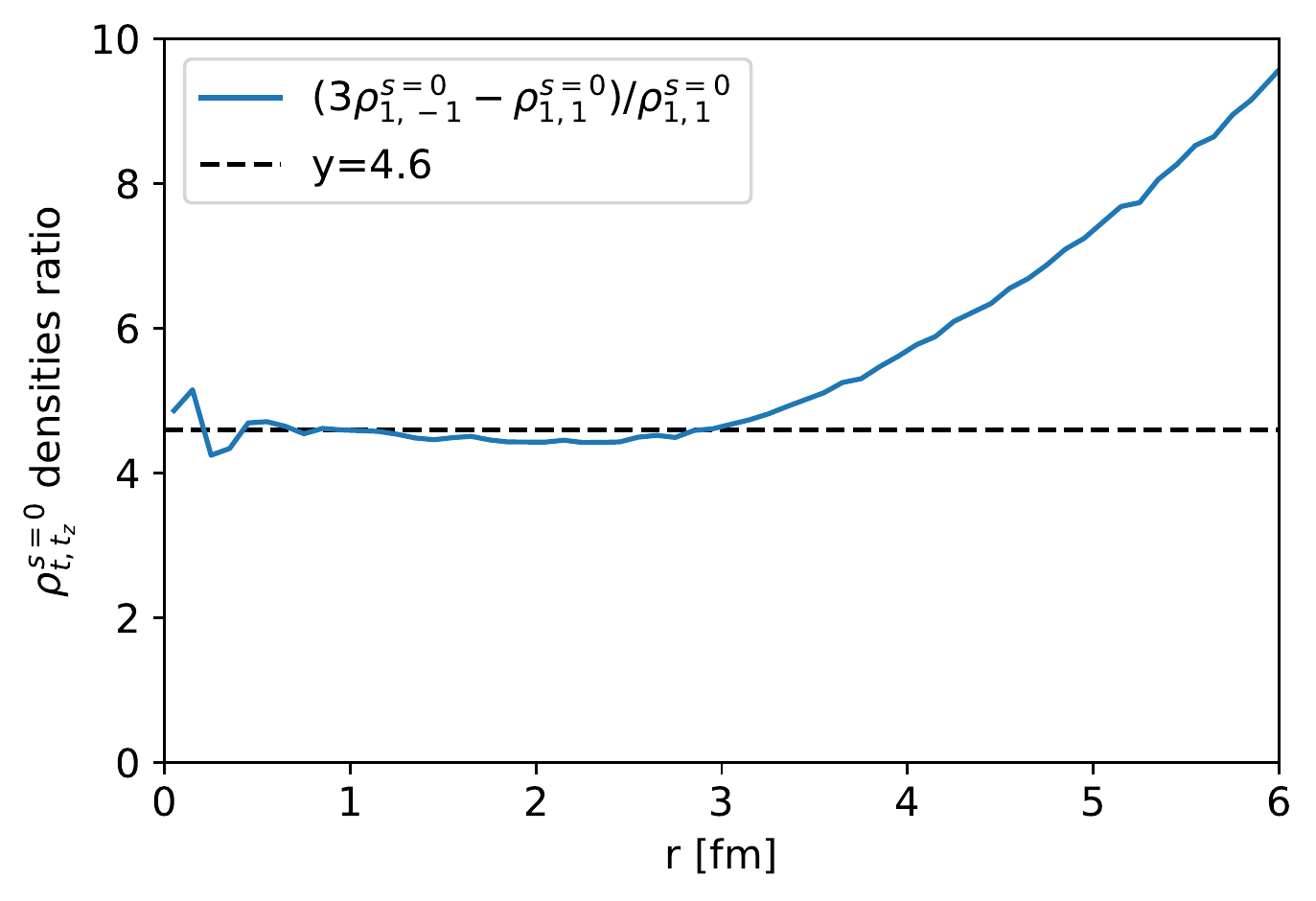}
\caption{\label{Li7_A-2_nn}
Ratio of VMC spin-zero two-body densities of $^7$Li ground state ($T=1/2,T_z=-1/2$)
using the AV18+UX potential, following Eq. \eqref{eq:T_A-2_ratio_nn}.
The horizontal dashed line shows the value of this ratio at short distances.
}
\end{center}\end{figure}

As an example, for $T=1/2$, $T_z=-1/2$ nucleus we get
\begin{align}
\langle \tilde{A}({1}/{2})| \tilde{A}({1}/{2}) \rangle
=
\frac{2}{A(A-1)}
\left[
\frac{3}{2}
C^{t=1}_{t_z=-1}
-
\frac{1}{2} C^{t=1}_{t_z=1}
\right]
\end{align}
\begin{align}
\langle \tilde{A}(3/2)| \tilde{A}(3/2) \rangle
=
\frac{2}{A(A-1)} 2 C^{t=1}_{t_z=1}.
\end{align}
Now, looking for example on a $t=1$, $t_z=-1$ ($nn$) pair
with small relative distance $r$
(inside a $T=1/2$, $T_z=-1/2$ nucleus),
we can ask what is the relative probability of finding the $A-2$ nucleon system in a state
with $T^{A-2}=1/2$ or $T^{A-2}=3/2$.
Based on Eq. \eqref{eq:C_t1_A_tilde} we obtain
\begin{align}
\frac{P_{nn}(T^{A-2}=1/2)}{P_{nn}(T^{A-2}=3/2)}
&=
\frac{|\langle \frac{1}{2} \frac{1}{2} 1 -1 | \frac{1}{2} -\frac{1}{2} \rangle|^2
\langle \tilde{A}(1/2)| \tilde{A}(1/2) \rangle}
{|\langle \frac{3}{2} \frac{1}{2} 1 -1 | \frac{1}{2} -\frac{1}{2} \rangle|^2
\langle \tilde{A}(3/2)| \tilde{A}(3/2) \rangle}
\nonumber \\ &=
\frac{ 3 C^{t=1}_{t_z=-1} -  C^{t=1}_{t_z=1} }
{C^{t=1}_{t_z=1}}.
\end{align}
Therefore, we can extract the relative contributions of 
$T^{A-2}=1/2$ and $T^{A-2}=3/2$ using only the $pp$ and $nn$ contact values.
Since the two-body densities at short distances are proportional to the contacts,
we can also express it directly using the two-body densities
\be \label{eq:T_A-2_ratio_nn}
\frac{P_{nn}(1/2)}{P_{nn}(3/2)}
=
\frac{ 3 \rho_{1,-1}^{s=0}(r\rightarrow 0) -  \rho_{1,1}^{s=0}(r\rightarrow 0) }
{\rho_{1,1}^{s=0}(r\rightarrow 0)}.
\ee
We used here the spin-zero densities because this is the spin
of the pair in the leading $t=1$ channel, although the spin-one
$t=1$ contribution is negligible at short distances and using the
total $\rho_{t,t_z}(r)$ densities leads to very similar results.
The last ratio is presented in Fig. \ref{Li7_A-2_nn} for the 
$^7$Li nucleus based on VMC calculations.
We can deduce that when a $nn$ pair is close together in $^7$Li, the $A-2$
spectators have approximately 4.6 larger probability to be in a $T=1/2$
state than a $T=3/2$ state.

This example was specifically for a $nn$ pair in a $T=1/2,T_z=-1/2$ nucleus, but
similar relations can be derived for any nucleus and any $t=1$ pair.
For $T\geq 1$ it will involve the three $t=1$ contacts and three
$T^{A-2}$ components. 
The contact values are needed to extract 
the relative probabilities of the $A-2$ nucleon isospin components.
We used here VMC calculations for this purpose, but contact values can also
be extracted, for example, from exclusive electron-scattering data where the knocked-out
pair is detected \cite{schmidt20}. Therefore, it can be applicable for heavy nuclei beyond the reach of ab-initio calculations. This also means that we can extract information about the $A-2$ nucleon system by only measuring the knocked-out pair and not the $A-2$ nucleons.

This information about the $T^{A-2}$ probabilities
is relevant for modeling the $A-2$ nucleon part of nuclei. It can be used,
for example, to estimate the maximum probability that the $A-2$ nucleon
subsystem is found in its ground state. It should also be useful for models of 
the spectral function, used to calculate lepton scattering cross sections, where the average excitation energy of the $A-2$ spectators is required \cite{Weiss:2018tbu,Andreoli:2021cxo}. This might also be relevant for the analysis of inverse-kinematics
scattering experiments, where the $A-2$ nucleon system is detected \cite{Patsyuk:2021fju}.
For example, if the probability of the A-2 nucleon system to be in different energy levels can be measured in such experiments using $\gamma$-ray detection, our predictions regarding the $T^{A-2}$ probabilities can be tested. Such studies will allow a further experimental examination of the short-distance wave function factorization to a single leading $t=1$ channel, which is required to obtain our predictions.

\section{Summary} \label{sec:summary}
In this work, we have derived general relations
concerning two-body distributions based on isospin symmetry.
These relations would be exact if isospin symmetry was exact in nuclear systems.
They were successfully tested against ab-initio VMC calculations with nuclear potential models that include realistic (and small) isospin-breaking terms.
Some of these relations connect isospin-one $pp$, $nn$, and $pn$ two-body densities (or momentum distributions)
in a given $T=0$ or $T=1/2$ nucleus. We also concluded that no such relations exist
for $T\geq1$ nuclei (aside from $T_z=0$ nuclei, where the $pp$ and $nn$
densities are equal). 
Another set of connections shows that, for any isospin multiplet, one can obtain
the two-body densities $\rho_{t,t_z}(r)$ (or the equivalent two-body momentum distributions)
of any nucleus in the multiplet if they are known for a single nucleus in this multiplet.
We have explicitly derived the relations for mirror nuclei and $T=1$ and $T=3/2$ multiplets. The relations for $T>3/2$ multiplets can be similarly obtained.
A connection between $0\nu\beta\beta$ decay densities of mirror nuclei and the two-body densities $\rho_{t,t_z}(r)$ was also revealed.

These results can be useful for guiding different models or estimations
that rely on two-body densities, which are needed for the calculation of two-body observables. 
They can also be utilized for benchmarking numerical codes and reducing 
the cost of calculations by calculating only independent quantities that are not related
via isospin symmetry (assuming isospin breaking effects are negligible).
In addition, isospin-breaking effects can be studied by looking for
violations of these relations.
As discussed before, they can also be used to obtain information regarding excited states
from ground-state nuclei in the same multiplet. 

Finally, we have focused on the implications of isospin symmetry on the study of nuclear SRCs. The relations discussed above lead to equivalent 
connections between different nuclear contacts,
that measure the number of SRC pairs in nuclei. These connections will allow a more accurate interpretation of ab-initio calculations and experimental data sensitive
to SRCs, and specifically a more accurate extraction of contact values for $T=1/2$ nuclei. This might be relevant for some of the upcoming experiments in the field, see e.g. Ref. \cite{Arrington_E12-06-105}. The $0\nu\beta\beta$ relations result in a connection between the relevant contact values, which reveals a connection between electron-scattering reactions and $\Delta T = 0$ $0\nu\beta\beta$ transitions.
Additionally, we have shown that isospin-symmetry arguments allow us to obtain information regarding the
isospin structure of the $A-2$ nucleon subsystem when an SRC pair is formed. The relative probability
of the different isospin components of that subsystem can be calculated using only the number of
SRC pairs in each channel. As mentioned above, this can be relevant for spectral-function models and inverse-kinematic experiments.

\section{Acknowledgement}
We would like to thank E. Piasetzky, J. Kahlbow and N. Barnea for their comments and suggestions. 
The present research is supported by the U.S. Department of Energy, Office of Science, Office of Nuclear Physics, under contract DE-AC02-06CH11357, the NUCLEI SciDAC program (A.L., R.B.W.), and by the Laboratory Directed Research and Development program of Los Alamos National Laboratory under project number 20210763PRD1 (R.W.).
The work of A.L. is also supported by the DOE Early Career Research Program award. Quantum Monte Carlo calculations were performed on the parallel computers of the Laboratory Computing Resource Center, Argonne National Laboratory, the computers of the Argonne Leadership Computing Facility via the INCITE grant ``Ab-initio nuclear structure and nuclear reactions'', and the 2020/2021 ALCC grant ``Chiral Nuclear Interactions from Nuclei to Nucleonic Matter''.


\appendix

\section{Isospin eigenstates} \label{sec:TTz}

We explain here in detail our claim that the functions given in Eqs. \eqref{eq:t1_TTz_func} and \eqref{eq:t0_TTz_func} are isospin eigenstates with the same $T,T_z$ quantum numbers as the full wave function $\Psi$. $\Psi$ obeys
\be
\hat{T}^2 | \Psi \ket = T(T+1) | \Psi \ket
\ee
and
\be
\hat{T}_z | \Psi \ket = T_z | \Psi \ket.
\ee
Starting with $\hat{T}_z$, we can multiply the last equation by $\varphi_m^{t\dagger}(\bs{r}_{12})$ and integrate over $\bs{r}_{12}$. $\varphi_m^{t\dagger}(\bs{r}_{12})$ is isospin independent [because the isospin part was explicitly separated, see Eq. \eqref{eq:basis}] so it can be exchanged with the operator $\hat{T}_z$ and we get
\be
\hat{T}_z \int d^3r_{12} \varphi_m^{t\dagger}(\bs{r}_{12}) \Psi(\bs{r}_1,...,\bs{r}_A) = T_z \int d^3r_{12} \varphi_m^{t\dagger}(\bs{r}_{12}) \Psi.
\ee
The basis states $\varphi_m^{t,t_z}$ are orthonormal. This directly means that for a given $t$ (and different values of $m$), the functions $\varphi_m^{t}$  are orthonormal as well because $\eta_{t,t_z}^\dagger \eta_{t,t_z} = 1$. Now consider two functions with different isospin, i.e. $\varphi_m^{t=0}$ and $\varphi_{m'}^{t'=1}$. Different isospin values dictate that either the spin or the parity of these two functions must also be different, due to the Pauli exclusion principle. Therefore, they are orthogonal as well. As a result, using the expansion of Eq. \eqref{eq:psi_expan}, we obtain
\be
\hat{T}_z \sum_{t_z} \eta_{t,t_z} S_m^{t,t_z}
=
T_z \sum_{t_z} \eta_{t,t_z} S_m^{t,t_z}.
\ee
An equivalent result can be similarly obtained for the operator $\hat{T}^2$, and so we conclude that the functions of the form $\sum_{t_z} \eta_{t,t_z} S_m^{t,t_z}$ have the same $T,T_z$ quantum numbers as $\Psi$. These are exactly the functions appearing in Eqs. \eqref{eq:t1_TTz_func} and \eqref{eq:t0_TTz_func}.

\section{$T=1/2$ relation}  \label{sec:Thalf_single_nucleus}

We provide here more details on the derivation of Eq. \eqref{eq:T_half}. Based on Eq. \eqref{eq:rho_t1} we obtain the following expressions for the $\rho_{1,t_z}(T=1/2,T_z=1/2)$ densities:
\be
\rho_{1,1}\left(\frac{1}{2},\frac{1}{2}\right) 
=
\frac{A(A-1)}{2} \frac{1}{4\pi r^2}
\left[ \frac{2}{3} D\left(\frac{1}{2}\right) + \frac{1}{6} D\left(\frac{3}{2}\right) \right]
\ee
\be
\rho_{1,0}\left(\frac{1}{2},\frac{1}{2}\right) 
=
\frac{A(A-1)}{2} \frac{1}{4\pi r^2}
\left[ \frac{1}{3} D\left(\frac{1}{2}\right) + \frac{1}{3} D\left(\frac{3}{2}\right) \right]
\ee
\be
\rho_{1,-1}\left(\frac{1}{2},\frac{1}{2}\right) 
=
\frac{A(A-1)}{2} \frac{1}{4\pi r^2}
\left[ \frac{1}{2} D\left(\frac{3}{2}\right) \right],
\ee
where we denote by $D(T^{A-2})$ the expression given in Eq. \eqref{eq:A-2_contr}. We can now directly see that indeed
\be
2 \rho_{1,0}\left(\frac{1}{2},\frac{1}{2}\right) = \rho_{1,1}\left(\frac{1}{2},\frac{1}{2}\right) + \rho_{1,-1}\left(\frac{1}{2},\frac{1}{2}\right), 
\ee
as argued in Eq. \eqref{eq:T_half}.
The same relation is obtained for $T_z=-1/2$.

\bibliography{biblio}

\end{document}